\documentclass[preprint2]{aastex}

\def\kms{\mbox{km/s}}
\def\pc{\mbox{pc}}
\def\kpc{\mbox{kpc}}

\def\Msun{\mbox{M$_\odot$}}
\def\Lsun{\mbox{L$_\odot$}}
\def\msun{\mbox{M$_\odot$}}

\def\Msunh{\mbox{$h^{-1}$M$_\odot$}}

\def\Mvir{\mbox{$M_{\rm vir}$}}
\def\LCDM{{\char'3CDM}}
\def\mathnew{\mathsurround=0pt}
\def\simov#1#2{\lower .5pt\vbox{\baselineskip0pt
    \lineskip-.5pt\ialign{$\mathnew#1\hfil##\hfil$\crcr#2\crcr\sim\crcr}}}

\def\'#1{\ifx#1i{\accent"13\i}\else{\accent"13#1}\fi}

\shorttitle{Milky Way galaxy}
\shortauthors{Klypin et al.}
\begin{document}
\title{\LCDM-based models for the Milky Way and M31 \\ I: Dynamical Models} 

\author{Anatoly Klypin}
\affil{Astronomy Department, New Mexico State University, Box 30001, Department
4500, Las Cruces, NM 88003-0001}
\author{HongSheng Zhao and Rachel S. Somerville}
\affil{Institute of Astronomy, Madingley Rd., Cambridge, CB3 0HA, UK}

\begin{abstract}
We apply standard disk formation theory with adiabatic contraction
within cuspy halo models predicted by the standard \LCDM\
cosmology. The resulting models are confronted with the broad range of
observational data available for the Milky Way and M31 galaxies. We
find that there is a narrow range of parameters which can satisfy the
observational constraints, but within this range, the models score
remarkably well. Our favored models have virial masses $10^{12}\,
\Msun$ and $1.6\times 10^{12}\, \Msun$ for the Galaxy and for M31
respectively, average spin parameters $\lambda \approx 0.03-0.05$, and
concentrations $C_{\rm vir}= 10-17$, typical for halos of this mass in
the standard \LCDM\ cosmology. The models require neither dark matter
modifications nor flat cores to fit the observational data. We explore
two types of models, with and without the exchange of angular momentum
between the dark matter and the baryons. The models without exchange
give reasonable rotation curves, fulfill constraints in the solar
neighborhood, and satisfy constraints at larger radii, but they may be
problematic for fast rotating central bars.  We explore models in
which the baryons experience additional contraction due to loss of
angular momentum to the surrounding dark matter (perhaps via a
bar-like mode).  These models produce similar global properties, but
the dark matter is only a fourth of the total mass in the central
3~kpc region, allowing a fast rotating bar to persist.  According to
preliminary calculations, our model galaxies probably have sufficient
baryonic mass in the central $\sim3.5$ kpc to reproduce recent
observational values of the optical depth to microlensing events
towards the Galactic center.  Our dynamical models unequivocally
require that about half of all the gas inside the virial radius must
not be in the disk or in the bulge, a result that is obtained
naturally in standard semi-analytic models. Assuming that the Milky
Way is ``typical'', we investigate whether the range of virial masses
allowed by our dynamical models is compatible with constraints from
the galaxy luminosity function. We find that if the Milky Way has a
luminosity $M_K =-24.0$, then these constraints are satisfied, but if
it is more luminous (as expected if it lies on the Tully-Fisher
relation), then the predicted space density is larger than the
observed space density of galaxies of the corresponding luminosity by
a factor of 1.5--2. We conclude that observed rotation curves and
dynamical properties of ``normal'' spiral galaxies appear to be
consistent with standard \LCDM.
\end{abstract}  

 
\section{Introduction}

Modeling the mass distribution within the Milky Way (MW) and M31
galaxies is a classical problem which has seen many past iterations
\citep{Einasto72,Innanen73,Schmidt75,Einasto79}. Because of our unique
ability to observe these systems in exquisite detail, this problem has
continued to be a fertile testing ground for theories of galactic
structure and galaxy formation \citep{Kent89,DehnenBinney98,
WilkinsonEvans, EvansWilkinson2000,OllingMerrifield01}. In the absence
of a coherent halo formation theory, early models were plagued by the
arbitrary structure parameters of the dark halos.  Recent developments
in the theory of dark matter halo formation make this sort of modeling
much more meaningful because the halo parameters (such as total mass,
peak rotation speed and concentration parameter) may no longer be
considered free parameters but are correlated, and these correlations
are specified within the context of a given Cold Dark Matter cosmology
\citep[e.g.][and references therein]{Bullocketal2001}.

While very successful on large scales, the standard cosmological
theory is experiencing mounting difficulties on the scales of galaxies
and dwarf galaxies. The so-called ``sub-structure'' problem (the
predicted over-abundance of sub-halos relative to observed dwarf
satellites of the MW and the M31 galaxies \citep{KKVP99, MooreSat})
may have been successfully resolved by invoking the presence of a
photoionizing background that ``squelches'' star-formation in small
halos \citep{BKW00,squelch,benson:01}. However, the so-called ``cusp''
and ``concentration'' problems remain thorny.  The most acute
manifestations of these problems are encountered for objects which are
dark matter dominated --- namely, the central parts of low surface
brightness galaxies and dwarf galaxies \citep{Moore94, FloresPrimack,
Moore99, vdb2000}.  For example, the high central density of \LCDM\
dark matter halos appears to be inconsistent with the observed
rotation curves of dwarf and low surface brightness galaxies
\citep{Blais-Ouellette99,Cote2000,Blais-Ouellette, deBlok01,alam:01}.

It is interesting to ask how \LCDM\ models fare in their predictions
for luminous, high surface brightness galaxies, such as the Milky Way
and M31.  These systems probe the dark matter properties on mass
scales of $10^{12}\, \Msun$, which are significantly larger than the
$10^8-10^9\, \Msun$ scales tested by inner rotation curves of dwarf
galaxies. The Milky Way is particularly interesting for constraining
the central cusp of the dark matter because the inner 3--10~kpc region
of the Milky Way is well studied.  Apart from the rotation curve and
integrated light distribution, the Milky Way offers many more
opportunities to study very detailed data which are not generally
available for extragalactic systems, e.g. individual stellar radial
velocities and proper motions in the solar neighbourhood, and
recently, Galactic bulge microlensing events.

It is likely that the halo of our Galaxy consists mostly of weakly
interacting elementary particles (wimps) rather than black holes or
failed stars (machos), a necessary condition to apply the standard
cosmological approach.  The very low microlensing counts by machos in
the inner 50~kpc halo of the MW imply that elementary particles make
up $\ge (80-90)\%$ of the halo
\citep{Alcock00,Lasserre00,ZhaoEvans00}.  Therefore it should be valid
to apply halo formation models to our Galaxy, which predict an outer
dark matter profile of the form $\rho \propto r^{-3}$. In this regard,
earlier halo models using cored isothermal halos with a $\rho \propto
r^{-2}$ outer profile and a divergent mass
\citep[e.g.][]{OllingMerrifield01}, or very steep outer profiles $\rho
\propto r^{-5}$ \citep[e.g.][]{EvansWilkinson2000} are unphysical and
should be updated.

\citet{nst2000} and \citet{Eke01} were the first to use mass modeling
of the Milky Way to constrain cosmological models.  The dark mass
inside the solar radius was compared with observational limits derived
from previous mass modeling.  This could give misleading results
because of inconsistent modeling of the halo profiles.  Also unlike in
galaxies with low baryon content such as dwarf or LSB galaxies, the
density of the inner halo can deviate significantly from the original
\citet[NFW;][]{nfw:96} profile in high-surface brightness galaxies
because of strong compression during the baryonic collapse
\citep{MoMaoWhite98}.  Given the coupling of the inner and outer parts
of the halo, it is necessary to repeat the mass modeling of the Milky
Way and M31 including a self-consistent treatment of the baryons and
dark matter component.

The amount of dark matter inside the central $\sim 3$~kpc of the Milky
Way and M31 introduces another layer of problems. It is very likely
that both galaxies have fast rotating bars. A rotating bar experiences
dynamical friction induced by the dark matter. If the mass of the dark
matter surrounding the bar is too large, the dynamical friction could
slow down the bar rotation over a short timescale \citep{Weinberg85,
Debattista00}, in which case, we should not expect to see fast rotating
bars. 
This argument sometimes is presented as the
most severe problem the \LCDM\ cosmological models encounter on small
scales \citep{Sellwood2000, SellwoodKosowski, Evans01, BE2001}. It is clear
that some effect of this kind must exist, but a realistic estimate of
its magnitude is still difficult for current analytic works
\citep{Weinberg85} and $N-$body simulations \citep{Debattista00}.

The COBE map of the Galaxy and the high rate of microlensing events
towards the Galactic bulge have been used as arguments that most of
the inner Galaxy's mass is in a heavy stellar bar
\citep{Zhao96,ZhaoMao96}.  This leaves very little room for a massive
dark matter halo \citep{BBG00}. We address these issues in this paper,
and will treat them in more detail in a paper in preparation (Zhao et
al. 2001).

Modeling the Milky Way and M31 turns out to be challenging because of
their complex formation processes.  The present-day dark matter halos
can have density profiles very different from the original dark matter
prediction (e.g. the NFW profile).  They are squeezed and deformed by
the gravitational force of the baryons, which collapse out of the NFW
halo to form heavy disks and dense nuclear bulges.  The amount of
squeezing also depends on the angular momentum transfer between the
baryons and the dark matter halo.  The dark matter cusp is ellusive to
detection because the gravitational force at the center is dominated
by the baryons.  Apart from detailed data, the Milky Way provides a
good laboratory for studying the transfer of angular momentum from the
rapidly rotating baryons to the dark matter halo through dynamical
friction.  Such a transfer must have taken place in the MW and M31
since both have a prominent triaxial bulge/bar and spiral arms in
their disks.  An analytic formulation for the angular momentum
transfer is given in an accompanying letter (Paper III).

Studying the MW halo can set general limits on dark matter halos only
if the MW is typical among galaxies that form in a dark matter halo of
a given mass. This ``typicality'' hypothesis can be tested to some
degree with our parallel study of the nearby galaxy M31. It can also
be tested by producing many realizations of a Milky Way-like galaxy
using a Monte Carlo-based semi-analytic model of galaxy formation.
The semi-analytic approach allows us to make statistical predictions
of the formation histories of ensembles of galaxies hosted by halos of
a given mass, and to incorporate the combined stochastic effects of
distributions of mass accretion and merging history, halo spin
parameter, and gas accretion and star formation history. This approach
has the unique ability to allow us to estimate the chances of finding
a certain combination of structural parameters that simultaneously
produce the observed stellar populations and gravitational potential
in galaxies like the MW and M31. We pursue this question in a
companion paper (Paper II).

In summary, the goal of this series of papers is to test whether the
current standard theory of galaxy formation, which seems to be very
successful on large scales and very problematic on dwarf galaxy
scales, can produce $L_*$ galaxies such as the MW and M31.  In this
paper, this is done by combining classical dynamical modeling of
detailed data for the MW and M31 with current cosmologically-based
models of dark halos and disk formation.

\section{Model Ingredients}
\label{sec:TheoConstraints}
\subsection{Mass components}
In traditional mass models of the MW one decomposes the baryons into
several descriptive components: the nucleus, the bulge, the bar, the
spheroid, the thin disk, the thick disk, and the cold interstellar
medium in the disk.  It is somewhat problematic to determine where a
component starts and ends, and this invariably adds to confusion of
terminology.  For example, the bar sits in the overlap region between
the bulge and the disk, and most stars in this region rotate in the
same sense with similar angular speed.  So is the bar a distinct
identity on top of an oblate bulge, is it merely the triaxial bulge,
or is it the non-axisymmetric part of the disk inside corotation?  The
lack of a 3D map of the Milky Way makes it uncertain whether or not
the disk is truncated inside the corotation of the bar. These
distinctions are subjective at some level and sometimes unnecessary
when the important quantity is the total mass distribution of the
baryons.

We take a different approach.  We divide up the mass of the Galaxy
into only two components: a dark halo component and a summed-up
baryonic component.  We describe this baryonic sum by a function
$M_b(<r)$, which is the combined mass of baryons inside the radius
$r$.  It is understood that the mass distribution is far from
spherical.  In fact it is highly flattened, and triaxial inside $\sim
3.5$ kpc.  We incorporate some minimal modeling of the dynamical
effects of flattening and triaxiality, but without repeating earlier
more rigorous 3-dimensional triaxial dynamical models
\citep{Zhao96,Hafner00}.

Our models for the bulge/bar are motivated by the \citet{Zhao96} model
of the galactic bulge/bar. That model used COBE DIRBE de-reddened
infrared maps \citep{Weiland94} and a collection of stellar kinematics
data in the direction of the central parts of the galactic bulge. The
fit to the data indicated that the bar has an elongated boxy triaxial
shape. \citet{Zhao96} added a steep oblate nucleus to the best-fit
so-called ``G2'' model of \citet{Dwek95}.  The nucleus and a massive
central black hole are important for modeling the compressed density
of the dark matter within 100 pc.  The total mass of the bulge, the
bar, and the nucleus was $(2.2\pm 0.2) \times 10^{10}\, \Msun$. No dark
matter was explicitly included in the model of \citet{Zhao96}.  A
massive Miyamo-Nagai analytic disk potential was used to make the
rotation curve flat within 8 kpc.  Here, instead of the Miyamo-Nagai
disk, we use a double-exponential disk as in \citet{Kent91}.  The
combined density of nucleus, bulge/bar, and disk are modeled by the
following three components:
\begin{eqnarray}
\label{eq:massmodela}
\rho_{\rm b} &=& \rho_1 + \rho_2 + \rho_3,\\
\rho_1 &=& \rho_{1,0} \left(s_1\right)^{-1.85}\!\!\exp\left(-s_1\right) ,\\
\rho_2 &=& \rho_{2,0} \exp\left(-s_2^2/2\right) ,\\
\rho_3 &=& \rho_{3,0} \exp\left(-s_3 \right),
\label{eq:massmodelz}
\end{eqnarray}
where $\rho_{1,0}$, $\rho_{2,0}$, and $\rho_{3,0}$ are characteristic
densities, determined by the corresponding total masses $M_1$, $M_2$, and
$M_3$ of the components $\rho_1$, $\rho_2$, and $\rho_3$.  
The dimensionless radii $s_1$, $s_2$, and $s_3$ are given by
\begin{eqnarray}
s_1^2 &=& {0.6^{2}(x^2+y^2)+z^2 \over z_0^2} \\
s_2^4 &=& {\left[\left(0.26x\right)^2 + \left(0.42y\right)^2\right]^2
+z^4 \over z_0^4},\\
s_3     &=& {\sqrt{x^2+y^2}+12 |z| \over r_d}.
\label{eq:bulge}
\end{eqnarray}
\noindent 
The above aspect ratios in the components $\rho_1(s1)$, $\rho_2(s2)$
and $\rho_3(s3)$ are taken from \citet{Zhao96} to reproduce the shape
of the COBE bar and the disk of \citet{Kent91}.  We take the vertical
scale height $z_0=400\pc$ from \citet{Zhao96}, and take the disk
scale-length to be $r_d=3000$ or $3500$~pc, and the distance to the
Galactic center to be $R_0=8000$ or $8500$~pc.  These parameter
combinations produce acceptable fits to the light distribution of the
Milky Way after proper normalization of the three
components. \citet{Zhao96} normalizes the $\rho_1$ and $\rho_2$
components so that they have equal strength at about a distance
$z=0.75z_0$ on the minor axis.  This fixes the ratio
$\rho_{1,0}:\rho_{2,0}$ so that $\rho_1+\rho_2$ joins smoothly from an
observed power-law nucleus to the observed COBE bar.  The ratio
$\rho_{2,0}:\rho_{3,0}$ is fixed so that $(M_1+M_2):M_3=1:5$,
approximately matching the bulge-to-disk K-band luminosity ratio
$1.1\times 10^{10}\Lsun:4.9\times 10^{10}\Lsun$ of \citet{Kent91}.
Kent et al. in fact introduced a linear tapering of their disk scale
length from 250 pc at the solar radius to 167 pc at the center for
better matching of the observed surface brightness map.  This makes
their disk less massive than our double-exponential disk if both are
normalized to the surface density and the volume density at the solar
radius.  The same is true of the triaxial models of
\citet{Freudenreich98}, where the axisymmetric disk is tailored into a
triaxial hole inside 3.5 kpc.

We do not introduce this fine tailoring of the disk, instead favoring
the mathematical simplicity of a $\rho_2$ bar and a $\rho_3$ disk,
rather than a true bar or disk.  The extra mass contained in our
$\rho_2$ disk is at the price of reducing the mass of our $\rho_3$ bar
given the same budget of mass or light within the central 3.5 kpc.

We have deliberately chosen not to label the three components
$\rho_1$, $\rho_2$ and $\rho_3$ as nucleus, bar and disk given the
uncertain truncations of these latter components.  Instead they should
be treated as convenient functions to describe the overall
distribution of baryons.  There is no separate thick disk component.
When computing the rotation curve we approximate $\rho_3$ as a
razor-thin disk, and $\rho_1$ and $\rho_2$ as spherical.  The enclosed
baryonic mass is computed by spherically averaging $\rho_b$: 
\begin{equation}\label{eq:Mb}
M_b(<r) = m_{BH} + 4\pi \!\!\int_0^r\!\! \left<\rho_b\right> r^2 dr \, ,
\end{equation}
where
\begin{equation}
\left<\rho_b\right> \equiv \int \!\! \rho_b {d\Omega \over 4\pi}
\end{equation}
A more user-friendly analytic approximation of $M_b(r)$ is:
\begin{eqnarray}
M_b (<r) = & m_{BH} + M_1 \left[ 1- \exp(-1.15x_1)\right] + \nonumber \\
           & M_2 \left[ {\rm erf}\left({x_2 \over \sqrt{2}}\right) - 
  \sqrt{{2 \over \pi}} x_2 \exp\left(-{x_2^2 \over 2}\right) \right] + \nonumber \\
           & M_3 \left[1-(1+x_3)\exp(-{x_3})\right],
\end{eqnarray}
where $x_1 \equiv r/r_1,\ x_2 \equiv r/r_2,\ x_3 \equiv r/r_3\,$ and
$r_1$ and $M_1$ are the scale-length and mass of the nucleus, $r_2$
and $M_2$ the bulge/bar, and $r_3$ and $M_3$ the disk.  We use
eq.~(\ref{eq:Mb})in our calculations.
The total mass of the Galaxy inside a radius $r$ is computed as
\begin{equation}
 M(r) =M_{\rm b}(r)+M_{\rm dm}(r),
\label{eq:adiabat2}
\end{equation}
where $M_{\rm dm}(r)$ is the mass profile of the dark matter.  We
include a central black hole of mass $m_{BH}=2.7 \times 10^6\, \Msun$
in the Milky Way models and $m_{BH}=3.7 \times 10^7\,\Msun$ in the M31
models.

In our Milky Way models the parameters of the bulge are kept close to
those of the original \citet{Zhao96} bulge. This is done to preserve
the agreement with the COBE data and with estimates of stellar radial
velocities. Another reason is the microlensing counts, which may be a
problem for models of the Milky Way \citep{ZhaoMao96}. The rate of
microlensing events was estimated for the \citet{Zhao96} bulge and was
found to be consistent with the constraints on the counts
\citep{Peale}.

\subsection{Dark matter halo properties}
A sparse set of observational data can generally be fit by a large
range of models, and sometimes unphysical models give excellent fits.
\citet{DehnenBinney98} give plenty of examples of this kind.  
To obtain physically interesting models, it is important to impose
constraints based on a physical theory of halo formation, and to
select models which are not only compatible with data but are also
physically well-motivated.  In this paper we use three results from
halo formation theory: the shape of the mass density profile, the halo
concentration-mass relationship, and the distribution of halo spin
parameter (angular momentum).

Throughout this paper, we assume that the Hubble constant is equal to
$H_0=\,70\kms/{\rm Mpc}$, the Universe is flat and the contribution of
matter (dark and baryonic) to the critical density is equal to
$\Omega_0 =0.3$.  We assume that initially the dark matter density
profile is described by an NFW profile \citep[NFW]{nfw97}:
\begin{eqnarray}
\label{eq:NFWa}
	\rho_{\rm halo}(r) &=& \frac{\rho_s}{x(1+x)^2}, \quad x = r/r_s \\
	M_{\rm halo}(r) &=& 4\pi\rho_sr_s^3f(x) \\ &=& M_{\rm vir}f(x)/f(C), \\
	f(x) &=& \ln(1+x) -\frac{x}{1+x},\\  C &=& r_{\rm vir}/r_s,\\
	M_{\rm vir} &=&\frac{4\pi}{3}\rho_{cr}\Omega_0\delta_{\rm th}r_{\rm vir}^3
\label{eq:NFWz}
\end{eqnarray}
\noindent where $C$ and $M_{\rm vir}$ are the halo concentration and
virial mass, and $r_{\rm vir}$ is the virial radius. In the above
equations, the parameter $\rho_{cr}$ is the critical density of the
Universe and $\delta_{\rm th}$ is the overdensity of a collapsed
object in the ``top-hat'' collapse model ($\delta_{\rm th}\approx 340$
for our cosmological model). Two independent parameters --- $C$ and
$M_{\rm vir}$ --- completely define all relevant halo properties.

For some halos the central slope may be even steeper (-1.5) than the
-1 slope in eq.~(\ref{eq:NFWa}) \citep[e.g., ][]{Moore98, Ghigna2000,
Klypin2001}. For radii larger than 0.5-1\% of the virial radius
(1-3~kpc for our galaxies) the difference between this profile and
eq.~(\ref{eq:NFWa}) is relatively small \citep{Klypin2001}. Baryonic
infall has a tendency to reduce this difference even more, as we will
discuss in section~\ref{sec:compression}. Because the baryonic
compression is very large and still quite uncertain and because of the
relatively small differences in the density profiles, we will not
consider profiles with a steeper -1.5 inner slope here.

While we have introduced the halo concentration $C$ as an independent
parameter, in practice it is found to be strongly correlated with the
virial mass \citep{nfw97,Bullocketal2001}. For the \LCDM\ model,
\citet{Bullocketal2001} give $C=15-3.3\log(M_{\rm
vir}/10^{12}\, \Msunh)$. Statistical ensembles of halos at fixed mass
show 40\% rms fluctuations of the concentration around this mean. For
$10^{12}\, \Msun$ halos this gives a range of $C=11-21$. Halos of this
mass may host galaxies of all types, not only the Sb galaxies
considered in this paper. One may argue that perhaps ellipticals,
lenticulars, and Sa's form in high concentration halos and low surface
brightness (LSB) galaxies are hosted by low concentration halos. Thus,
the range for ``normal'' Sb galaxies may be somewhat narrower than the
full spread for all halos. Tentatively, we assume that the
concentration should be in the range $C=10-17$.

The angular momentum of dark matter halos is also theoretically
constrained. Halos in cosmological simulations have small amounts of
angular momentum, characterized by the dimensionless spin parameter
\begin{equation}
\lambda = J|E|^{1/2}/GM_{\rm vir}^{5/2}
\end{equation}
where $J$ is the angular momentum and $E$ is the total energy of the
halo. The spin parameter has a log-normal distribution, which does not
depend (or has very little dependence) on cosmological parameters,
halo mass, or redshift \citep{BarnesEfstathiou, LemsonKauffmann,
Vitvitska2001}.  We use the parameters given by
\citet{Vitvitska2001}. The maximum of the distribution of $\lambda$ is
at $\lambda=0.035$ and $\lambda$ has a 90\% probability of being in
the range
\begin{equation}
\lambda=0.02-0.10 \, .
\end{equation} 
Just as in the case of the concentration, it is likely that there is a
correlation of morphological type and spin, with Sb galaxies avoiding
halos with both extremely low and high values of $\lambda$.  

For each set of model assumptions, we can estimate the spin parameter
of its pre-collapse halo. After the fitting is done and we have the
rotation curve, we find the angular momentum of the disk: $J_{\rm
disk}=$$2\pi\int V_{\rm rot}\Sigma(r)r^2dr$, where $\Sigma(r)$ is the
surface density of the disk. We then assume that the specific angular
momentum of the whole system $j=J/M_{\rm vir}$ is equal to the
specific angular momentum of the present-day disk $j_{\rm disk}=J_{\rm
disk}/M_{\rm disk}$. This gives us the total angular momentum
$J$. Using the parameters of the pre-collapse halo we estimate the
potential energy of the system and, assuming virial equilibrium, find
the spin parameter. The assumption of virial equilibrium should be
quite reasonable because our systems are not interacting or merging.

The assumption that $j=j_{\rm disk}$ is a sensible and conventional
starting point, and should require little further discussion. We
neglect the angular momentum 
of the bulge in our calculations. This does not give large errors
because the angular momentum of bulges in our models is much smaller
that the angular momentum of disks. Because we restrict our modelling
to late type galaxies, bulge masses are also small ($<20$\% of the
disk mass), so in the worst case we are making 20\% error.  The
original angular momentum of the dark matter halo comes from
gravitational (tidal) interactions with its environment. Thus, the
dark matter and the gas experience the same torque in the process of
halo assembly and should initially have (almost) the same specific
angular momentum.  The main uncertainty is what happens to the gas and
its angular momentum from the moment it crosses the virial radius
until it finally settles into the disk. As we shall find, a
significant fraction of the gas must not end up in the disk. In our
models, about half of the gas is not found in the disk or bulge. The
models presented here do not address the question of where that gas
is; whether it is still inside the virial radius, was expelled from
the halo or was never in the halo. For simplicity, we assume that the
specific angular momentum was not affected by whatever processes
(presumably some form of feedback) were responsible for removing the
gas from the disk or preventing its collapse. Another issue is the
exchange of angular momentum between the baryons and the dark
matter. Some redistribution of the angular momentum may have happened,
but we do not know how significant it was. We consider two types of
models, with and without exchange of angular momentum. In both cases,
within a specific set of assumptions about the angular momentum
exchange (or lack thereof), we can thus find the initial angular
momentum of the gas which is presently in the disk.

\subsection{Baryonic compression}
\label{sec:baryons}
We start with models with no exchange of angular momentum between
different components.  The effects of baryonic infall are treated
following same approach as that of \citet{blumenthal}, \citet{flores},
and \citet{MoMaoWhite98}.  It is assumed that the compression occurs
adiabatically and the angular momentum of each component is preserved
as the baryons settle into the exponential disk. Assuming additionally
that the velocity anisotropy is preserved, one obtains the following
relation:
\begin{equation}
G \left[M_{\rm b}(r)+M_{\rm dm}(r) \right] r = G M_{\rm halo}(r_i)r_i,~
\label{eq:adiabat1}
\end{equation}
\noindent where $j$ is the specific angular momentum,
$r_i$ is the average radius of a dark matter particle
before the baryonic compression, $M_b(r)$ and $M_{\rm dm}$ are the
baryonic mass and dark matter mass inside the final radius $r$ and the
$M_{\rm halo}(r_i)$ is the halo mass before contraction, given by
\begin{equation}\label{eq:adiabat2b}
M_{\rm halo}(r_i) = {M_{\rm dm}(r) (\Omega_b+\Omega_{\rm dm})
\over \Omega_{\rm dm}}.
\end{equation}
Equation (\ref{eq:adiabat1}) can be considered as an equation for the
initial radius $r_i$ corresponding to a given final radius $r$. This
equation is solved numerically.

\subsection{Exchange of angular momentum between baryons and DM}

Models with exchange of angular momentum between the baryons and dark
matter are more complicated. The exchange probably happens at late
stages of the baryonic infall when the baryon density becomes large
and a non-axisymmetric component may develop due to the excitation of
spiral waves and/or bar-like modes.
Giant molecular clouds may play some role in this process.
Dynamical friction can then result in a transfer of angular momentum
from the baryons to the dark matter.  Because the dark matter gains
angular momentum, it moves further from the galactic center. Thus, the
density of the dark matter in the central region decreases.  In the
early stages of galaxy formation, when most of the baryons were still
in gaseous form, we might expect that those non-axisymmetric features
were more prevalent and more dynamically important then at the present
time when most of the baryons are locked in stars.

It is difficult to estimate the exact amount of angular momentum that
would be lost by the baryons in such a situation, but we know that it
cannot be very large. It is constrained by two factors: the angular
momentum of the disk at present, and the condition that the initial
spin parameter should be close to the typical (median) value for
halos. Models without the exchange of angular momentum give
$\lambda\approx 0.02-0.03$, which is already close to the most
probable spin $\lambda=0.035$ of a dark matter halo. If the disk loses
some angular momentum during its assembly, the estimate of the initial
spin parameter increases almost linearly with the lost angular
momentum.  If too much angular momentum is lost, the spin parameter
becomes unacceptably large and we would be forced to conclude that our
Galaxy formed from the extreme tail of the distribution.  Still,
because the distribution of spin parameter is quite broad, there is
some room for angular momentum exchange. The spin parameter can be two
times larger than for the no-exchange models and still be considered
``typical''. Thus for our model with angular momentum exchange, we
assume that the disk loses a factor of 1.5--2 of its angular momentum
during the collapse.  For simplicity we assume that the formation of
the disk happens in two stages. During the first stage, when the
baryons experience most of the collapse, they preserve their angular
momentum. During the second stage the disk shrinks further, losing
some of its angular momentum to the dark matter.

We use the approach outlined in the previous section to compute the
state of the system at the end of the first stage of adiabatic
compression (in which angular momentum is conserved).  We then
consider a spherical shell of dark matter with radius $r$, thickness
$dr$, density $\rho_{\rm dm}$, and specific angular momentum
\begin{equation}
j=r V_c = \sqrt{G\left[M_{\rm b}(r)+M_{\rm dm}(r) \right] r}.
\label{eq:jr}
\end{equation}
It has a total mass $dM_{\rm dm}$, and total angular momentum of 
\begin{eqnarray}
dJ & = & j \, dM_{\rm dm} \\
dM_{\rm dm}& = & 4\pi\rho_{{\rm dm}}r^2 dr\, .
\label{eq:dJ}
\end{eqnarray}

We estimate the amount of angular momentum lost by the baryons (disk
and bulge) when a mass element of baryons $\Delta M_b$ moves from
radius $r+dr$ to radius $r$. The baryons lose angular momentum
\begin{equation}
dJ_{\rm b}= dM_{\rm b}\left[(V_c+\frac{dV_c}{dr}dr)(r+dr)-V_cr\right].
\label{eq:difangularA}
\end{equation}
This angular momentum is deposited into the dark matter, which
produced the dynamical friction. Thus, the final angular momentum of the DM shell is
\begin{equation}
dJ_f= dJ + dJ_b. 
\label{eq:difangularB}
\end{equation}

Shells at different distances should be affected because the dynamical
friction is not a local process. Nevertheless, for simplicity, we
deposit all the angular momentum into the shell through which the disk
or bulge element moves. This is a reasonable approximation because (i)
a significant fraction of the dynamical friction is due to the
elements of dark matter closest to the sinking baryonic material, and
(ii) the volume affected by the deposition of the angular momentum for
all shells is very extended --- up to 20~kpc.

A shell of dark matter at radius $r$, which acquired specific
angular momentum 
\begin{equation}
j_f - j = dJ_b/dM_{\rm dm},
\label{eq:jfj}
\end{equation}
then moves to a new radius $r_f$. Here $j_f$ is the final specific angular
momentum and is related to the mass model by
\begin{equation}
j_f^2 = GM(r_f) r_f.
\label{eq:jf2}
\end{equation}
Equating initial and final angular momenta for each shell, we
get an implicit equation for the final radius $r_f$:
\begin{eqnarray}
\label{eq:exchangeangularA}
j_f &=& j \left[1+\frac{A\Delta M}
                  {4\pi\rho_{\rm dm}r^3}\right],\phantom{mmm} \\
A &=& 1+\frac{r}{V_c}\frac{dV_c}{dr},\\
\Delta M &=&M_{\rm b,f}-M_{\rm b}.
\label{eq:exchangeangularB}
\end{eqnarray}
Here $M=M_{\rm dm}+M_{\rm b}$ is the total mass inside a radius $r$.
Eq.~(\ref{eq:exchangeangularA}) is solved numerically. The solution
also gives the mass inside a final radius $r_f$.
Eq.~(\ref{eq:exchangeangularA}) has the same structure as
eq.~(\ref{eq:adiabat1}). The only difference is the term on the
right-hand-side, which is the correction due to angular momentum
deposition.

We can get a rough estimate of the effect if we neglect the term $A$
(which is close to unity) and introduce the average density excess
produced by infalling baryons $\Delta\rho=\Delta M/(4\pi r_i^3/3)$.
If we further assume that $M\propto r$ (a good approximation for most
radii), then eq.~(\ref{eq:exchangeangularA}) takes the form:
\begin{equation}
   r_f \approx r_i\left(1+\frac{\Delta\rho}{3\rho_{\rm dm}}\right)
\label{eq:effectang}
\end{equation}
It is clear that during the initial stages of the collapse, when the
density of the baryons was about ten times smaller than the density of
the dark matter, the exchange of angular momentum had little impact on
the dark matter: $r_f\approx r_i$. The effect peaks at around
$\Delta\rho =3\rho_{\rm dm}$. At even larger values of the density
ratio, the approximation fails because a small amount of dark matter
cannot exert significant dynamical friction on large mass of
baryons. Nevertheless, at the peak of its importance, the effect is
potentially quite large, with $r_f\approx 2r_i$, resulting in a
decrease of the dark matter density by a factor of ten.
 
\section{Observational Constraints}
\label{sec:ObsConstraints}
\subsection{Constraints for the Milky Way}
In this section we describe the observational data that we use to
confront our models.

Satellite dynamics and modeling of the Magellanic Clouds
\citep[e.g.][]{Zaritsky89, FichTremaine91, Lin95, Kochanek96} provide constraints
on the mass of our galaxy on large scales ($\simeq 50-100$ kpc).  The
mass of the Milky Way inside 100~kpc is estimated to be $(5.5\pm
1)\times 10^{11}\, \Msun$ from dynamics of the Magellanic Clouds,
\citep{Lin95}.  Using constraints from the escape velocity and motions
of satellite galaxies, \citet{Kochanek96} estimates the mass of the
Galaxy inside 100~kpc to be $(5-8) \times 10^{11}\, \Msun$.  We adopt
the following constraints on the mass inside $100\kpc$
\citep{DehnenBinney98}:
\begin{equation}
M_{R<100\kpc} =(7\pm 2.5)\times 10^{11}\, \Msun\, .
\end{equation}

For the outer part of the rotation curve of our Galaxy we use the data
summarized in \citet{DehnenBinney98}. The radial circular velocity
$v_r$ of an object at galactocentric coordinates $l$ and $b$ is
related to the circular velocity at radius r, $v_c(r)$, by
\begin{equation}
W(r/R_0) =\frac{v_r}{\sin l \cos b} =\frac{R_0}{r}v_c(r)-v_c(R_0),
\label{eq:WofR}
\end{equation}
\noindent where $R_0$ is the Sun's distance to the galactic center.
Data for $H_{\rm II}$ regions \citep{BrandBlitz93} and for classical
Cepheids \citep{Pont97} are used to derive $W(r/R_0)$ for up to twice
the solar distance.

The surface density of gas and stellar components at the solar radius
is estimated by \citet{KuijkenGilmore89} to be:
\begin{equation}
\Sigma_{\rm stars+gas} =48\pm 8 \, \Msun \pc^{-2}\label{eq:surface}
\end{equation}
We also use another local constraint: the vertical force $K_z$ at
1.1~kpc above the galactic plane. Using data on the kinematics of K
dwarfs, \citep{KuijkenGilmore89,KuijkenGilmore91} give the limit on
the total density of matter inside 1.1~kpc:
\begin{equation}
|K_z(R_0,z=1.1\kpc)|/2\pi G =(71\pm 6)\, \Msun\pc^{-2}
\end{equation}

Oort's constants: 
\begin{eqnarray}
A & = & (v_c/R-dv_c/dR)/2\\ 
B & = & -(v_c/R+dv_c/dR)/2
\end{eqnarray}
provide constraints on the circular velocity curve at the solar
radius. We adopt the same values as \citet{DehnenBinney98}:
\begin{eqnarray}
A & = & \phantom{-}14.5 \pm 1.5~{\rm km~ sec}^{-1}~{\rm kpc}^{-1} \\ 
B & = & -12.5 \pm 2~{\rm km~ sec}^{-1}~{\rm kpc}^{-1} \\
A - B & = & \phantom{-}27 \pm 1.5 ~{\rm km~ sec}^{-1}~{\rm kpc}^{-1}.
\label{eq:Oort}
\end{eqnarray}
The corresponding $v_c=(A-B)R_0$ at the solar neighborhood is
$200-240\kms$.

We use observations of terminal velocities $v_{\rm terminal}$ to
constrain the mass distribution and the rotational velocities inside
the solar radius. If $l$ is the galactic longitude, then for an
axisymmetric model the terminal velocity is related to the circular
velocity $v_c$ by
\begin{equation}
v_{\rm terminal} = v_c(R_0\sin l) - v_c(R_0)\sin l \label{eq:terminalV},
\end{equation}
\noindent where $R_0$ is the distance to the galactic center.  We make
use of data from the HI surveys of \citet{Knapp85} and
\citet{Kerr86}. The terminal velocities are compatible with the
results of \citet{Malhotra95}.  \citet{DehnenBinney98} present a more
comprehensive comparison of different surveys.

To constrain the mass in the central parts of the MW, we use data on
stellar motions.  We make use of the results of \citet{Genzel2000},
who found $M=3\times 10^7\, \Msun$ for the mass inside 10~pc and
$M=10^7\, \Msun$ at 4~pc.  Kinematics of OH/IR stars
\citep{Lindqvist1992} constrain the mass on the $\approx 100$~pc
scale.

More indirect constraints on the mass of baryons and dark matter in
the central part of the galaxy can be obtained from the existence of
persistent rapidly rotating bars and from the observed optical depth
to microlensing events in the direction of the Galactic center. Based
on the former, the work of \citet{Debattista00} suggests that dark
matter must make up less than a quarter of the dynamical mass within
one disk scale length. However, the halo models used in that work were
rather unphysical $n=3-5$ polytropic cored models, and so it is
difficult to know whether these results apply to cuspy CDM halos. From
the microlensing constraints, \citet{ZhaoMao96} concluded that all of
the dynamical mass within 3.5 kpc must be in an optimal stellar bar to
account for observed optical depth of $(3-10)\times 10^{-6}$. More
recent data suggest that the optical depth may be much lower than
this. Both of these pieces of evidence suggest that models with a
minimal dark matter contribution in the inner part are to be
preferred. We discuss how our models fare with respect to both of
these constraints in Section~\ref{sec:microlensing} and
\ref{sec:conclusions}; however, we do not impose either of these
constraints {\it a priori}.

We use the K-band luminosity of our Galaxy derived by
\citet{DrimmelSpergel2001}:
\begin{equation}
L_K =8.9\times 10^{10}L_{\odot}
\end{equation}
corresponding to a magnitude of $M_K = -24.0$.  This value is
consistent with the estimate of \citet{malhotra96}, but it is larger
than the value $6\times 10^{10}L_{\odot}$ adopted by
\citet{Kent91}. It is about 0.4--0.6 magnitudes fainter than the value
implied by the K-band Tully-Fisher relation of \citet{tp00}, assuming
a rotation velocity of 220 km/s for the Galaxy (see the discussion in
Section~\ref{sec:lf}).

\subsection{M31: light and kinematics}
\label{sec:obsM31}
Throughout this paper, we assume that the distance to M31 is 770~kpc,
based on Cepheids and red clump stars
\citep{FreedmanMadore,Kennicutt98,StanekGarnavich}. In some earlier
papers a smaller distance of 690~kpc was used. Results used below were
rescaled to our adopted distance. 

The M31 galaxy has a more extended disk than our galaxy. \citet{WK88}
studied the surface brightness distribution of M31 in four colors, U,
B, V, and R, tracing the disk to 20~kpc. The exponential length in the
R-band was found to be $5.7\pm 0.3~\kpc$, as compared with
$2.5-3.5~\kpc$ for our Galaxy. The scale length was found to be larger
in bluer bands ($6.4~\kpc$ in B). We use the R-band value in our
modeling because it is less affected by dust extinction and more
likely to trace the overall stellar mass distribution.  A significant
fraction of the light (1/3 to 1/2, depending on the band) is from the
bulge.  The bulge of M31 is triaxial, as evidenced by a significant
twist in isophotes and complex gas velocity patterns.  Like the Milky
Way, M31 has a nucleus as well.  In fact, the nucleus can be resolved
into two central peaks separated by about 1 pc.  A black hole of mass
$\sim 3.7\times10^7\, \Msun$ is harbored in the less luminous peak
(P2).

We model the bulge of M31 with a scaled-up version of the Milky Way
bulge (eqs.~\ref{eq:massmodela}-\ref{eq:massmodelz}) and disk, using a
disk scale radius of $r_d=5.7$kpc. We include a black hole with mass
$3.7\times10^7\, \Msun$ --- more than ten times more massive than the
Milky Way black hole.  R-band photometry along the major axis of M31
is taken from \citet{WK87}.  We also use $r-$band photometry from
\citet{Kent87}. The latter results are shifted up by 0.45~mag to match
the R-band surface brightness.
We apply a correction of 0.25~mag to both the bulge and disk to
account for extinction in our Galaxy, and apply an additional
correction of 0.74~mag to the disk for internal extinction
\citep{Kent89}.

We use the CO observations of \citet{Loinard95} to determine the
rotational velocity inside the central 10~kpc. HI measurements extend
the rotation curve to 30~kpc \citep{BrinksBurton84}. The CO and HI
velocities agree in the overlapping central region of the galaxy.  The
central 2~kpc region of M31 shows anomalously high velocities and a
misalignment of the major axes of the bulge and disk. Those anomalies
are attributed to the presence of a triaxial bar/bulge in M31
\citep[e.g.][]{Stark77, StarkBinney94, Berman2001}. Just as in the
case of the Milky Way, we use a simplified spherical model of the
central bulge.  The model can not reproduce the details of the
circular velocity function in the central 2-5~kpc, but it gives useful
constraints on the mass models.

Additional constraints are obtained from stellar kinematics in the
central region of M31. We use the estimates of the rotation velocity
$v_{\rm rot}$ and the line-of-sight velocity dispersion $\sigma$ from
\citet[Table 2]{KormendyBender}. We then estimate the mass, assuming
an isothermal distribution with isotropic velocities $M = 2(\sigma^2 +
v_{\rm rot}^2)r/G$. At distances 25$^{''}$-50$^{''}$ ($\approx
100-200$~pc) from the center of M31 the rotational velocity is small,
the velocity dispersion is almost constant (150\,\kms -160\,\kms) and the
assumed density profile of the nucleus is close to $r^{-2}$. Thus, our
approximation seems to be quite reasonable for these radii.

\section{Models of the Milky Way and M31}

\subsection{Comparison with Observations}
Our goal is to find whether a set of models, which we consider
plausible or theoretically well-motivated, are compatible with the
observational constraints. We do not attempt to find best-fit models,
but present a range of models and discuss whether they satisfy the
constraints we have adopted.  Once we find by trial and error that a
model is compatible with the constraints, we do not attempt to further
improve the quality of the fit by fine-tuning the parameters of the
model.

\begin{deluxetable}{llllll} 
\tablecolumns{12} 
\tablewidth{0pc} 
\tablecaption{The Milky Way galaxy: No exchange of angular momentum} 
\tablehead{ 
\colhead{\small Parameter} & \colhead{Constraints} & \colhead{Model $A_1$}   & \colhead{Model $A_2$}    & \colhead{Model $A_3$} & \colhead{Model $A_4$} \\
\colhead{} & & \colhead{Favored}   & \colhead{Max. Disk}    & \colhead{Max. C} & \colhead{Max. halo} \\
}
\startdata 
Virial mass $M_{\rm vir}$ (\Msun) & -- &	$1.0\times 10^{12}$ & $0.71\times 10^{12}$ &$1.0\times 10^{12}$ &$2.0\times 10^{12}$\\
Virial radius $r_{\rm vir}$ (\kpc) & -- & 258	 & 230 &258 &325\\
Halo concentration $C$ & 10-17 &			12 		& 5	 & 	17&10\\
Disk mass $M_3$ (eq.~(3); \Msun) & --	&	$4\times 10^{10}$ & $6\times 10^{10}$ & $3.5\times 10^{10}$ &$4.0\times 10^{10}$\\
Mass $M_1+M_2$ (eq.~(3), \Msun) & --	&	$0.8\times 10^{10}$ & $1.2\times 10^{10}$ &$0.7\times 10^{10}$ &$0.8\times 10^{10}$\\
Stellar mass $r<3.5$~kpc (\Msun) & --	&	$1.9\times 10^{10}$ & $3.1\times 10^{10}$ &$1.5\times 10^{10}$ &$1.9\times 10^{10}$\\
Disk exponential scale length  & 2.5-3.5 &	3.5 & 3.0 & 3.5 &3.5\\
\qquad  $r_d$ (\kpc) & 	 &  &  & \\
Maximum circular velocity  & -- & 228 & 246 & 235 & 237 \\
\qquad $V_{c, max}$ (\kms) & 	 &  &  & \\
Max. halo circular velocity & -- &  163 & 123 & 178 &197\\
\qquad no compression (\kms) &		 & & \\
Solar distance, $R_{\odot}$ (\kpc)& 7-8.5 &  8.0 & 8.5 & 8.5 & 8.5 \\ 
Baryon surface density at $R_{\odot}$,  &$48\pm8$ &53 & 62 & 40& 46\\
\qquad (\Msun\pc$^{-2}$)  & & & & \\
Total surf. density within 1.1~kpc  &$71\pm6$ &75 & 74 &63 & 69\\
\qquad at $R_{\odot}$ ($\Msun\pc^{-2}$)  & & & & \\
Oort's constants $A-B$ &$27\pm 1.5$ &26.8  &28.1 & 27.6& 27.9\\
\qquad   (km~s$^{-1}$~kpc$^{-1}$) & 	 &  &  & \\
Mass inside 100~\kpc~ ($10^{11}$ \Msun)  & $7.5\pm 2.5$	&$5.8$ &$3.8$ &$6.1$ &$9.0$ \\ 
Disk+bulge mass-to-light ratio &	-- &0.54		& 0.81 &0.47 &0.54\\
\qquad   $(M/L)_K$ (\Msun/\Lsun) & 	 &  &  & \\
Spin parameter $\lambda$ &0.02--0.10&	0.031 &0.022	&0.037 &0.018\\
$M_{\rm dm}/M_{\rm disk+bulge}(r<3{\rm kpc})$ & -- &0.95	 &0.40	&1.31 &1.02\\
Fraction of ``galactic'' baryons & -- &0.48	 &1.00	&0.42 &0.24\\
\enddata 
\end{deluxetable}

Tables 1 -- 3 list the parameters of our models. Most of the
parameters and values have already been introduced and discussed in
sections \ref{sec:TheoConstraints} and \ref{sec:ObsConstraints}.  The
last row in each table gives the ratio of the sum of disk and bulge
masses to the mass of baryons expected inside the virial radius ($f_b
M_{\rm vir}$). We assume the universal ratio of baryons to dark matter
is $f_b \equiv \Omega_b/(\Omega_{\rm dm} +\Omega_b) =0.1$. Our results
can be easily adjusted to any other ratio.

With the exception of models $A_2$ and $B_2$, all the models pass the
constraints we have adopted. Typically the fitting was done by first
assuming a mass and concentration for the halo and then by tuning the
disk parameters.

\begin{deluxetable}{llllll} 
\tablecolumns{12} 
\tablewidth{0pc} 
\tablecaption{The Milky Way galaxy: Models with the exchange of angular momentum} 
\tablehead{ 
\colhead{\small Parameter} & \colhead{Constraints} & \colhead{Model $B_1$}   & \colhead{Model $B_2$}    & \colhead{Model $B_3$} & \colhead{Model $B_4$} \\
\colhead{} & & \colhead{Favored}   & \colhead{Max. Disk}    & \colhead{Large Mass} & \colhead{Small Exch.} \\
}
\startdata 
Virial mass $M_{\rm vir}$ (\Msun) & -- &	$1.0\times 10^{12}$ & $0.71\times 10^{12}$ &$1.5\times 10^{12}$ &$1.0\times 10^{12}$\\
Virial radius $r_{\rm vir}$ (\kpc) & -- &	258 &230  & 295 & 258\\
Halo concentration $C$ & 10-17 &			12 		& 10	 & 	10&12\\
Disk mass $M_3$ (\Msun) & --	&	$5\times 10^{10}$ & $6\times 10^{10}$ & $5\times 10^{10}$ &$5\times 10^{10}$\\
Mass $M_1+M_2$ (\Msun) & --	&	$1\times 10^{10}$ & $1.2\times 10^{10}$ &$1\times 10^{10}$ &$1\times 10^{10}$\\
Stellar mass $r<3.5$~kpc (\Msun) & --	&	$2.7\times 10^{10}$ & $2.8\times 10^{10}$ &$2.7\times 10^{10}$ &$2.6\times 10^{10}$\\
Disk exponential scale length  & 2.5-3.5 &	3.0 & 3.5 & 3.0 &3.0\\
\qquad  $r_d$ (\kpc) & 	 &  &  & \\
Intermediate exponential scale  & -- &	6.0 & 7.0 & 6.0 &4.5\\
\qquad  $r_d$ (\kpc) & 	 &  &  & \\
Maximum circular velocity  & -- & $223$ & $216$ & 225 & 234 \\
\qquad $V_{c, max}$ (\kms) & 	 &  &  & \\
Max. halo circular velocity & -- &  163 & 139 & 178 &163\\
\qquad no infall (\kms) &		 & & \\
Solar distance $R_{\odot}$ (\kpc), & 7-8.5 &  8.5 & 8.5 & 8.5 & 8.5 \\ 
Baryon surface density at $R_{\odot}$,  &$48\pm8$ &52 & 69 & 52& 52\\
\qquad ($\Msun\pc^{-2}$)  & & & & \\
Total surf. density within 1.1~kpc  &$71\pm6$ &68 & 79 &70 &72 \\
\qquad at $R_{\odot}$ ($\Msun\pc^{-2})$  & & & & \\
Oort's constants $A-B$, &$27\pm 1.5$ &26.5  &25.4 & 26.5& 28.0\\
\qquad  (km~s$^{-1}$~kpc$^{-1}$) & 	 &  &  & \\
Mass inside 100~\kpc~,  & $7.5\pm 2.5$	&$6.0$ &$4.6$ &$7.7$ &$6.0$ \\ 
\qquad  ($10^{11}$\Msun) 	 &  &  & \\
Bulge+disk mass-to-light ratio &	-- &0.67		& 0.81 &0.67 &0.67\\
\qquad   $(M/L)_K$ (\Msun/\Lsun) & 	 &  &  & \\
Spin parameter $\lambda$ &0.02--0.10&	0.058 &0.076	&0.044 &0.050\\
$M_{\rm dm}/M_{\rm disk+bulge}(r<3{\rm kpc})$ & -- &0.24	 &0.14	&0.24 &0.31\\
Fraction of ``galactic'' baryons & -- &0.6	 &1.0	&0.4 &0.6\\
\enddata 
\end{deluxetable} 

\begin{deluxetable}{lll} 
\tablecolumns{11} 
\tablewidth{0pc} 
\tablecaption{Parameters of Models for the M31 galaxy} 
\tablehead{ 
\colhead{\small Parameter} & \colhead{Model $C_1$}  & \colhead{Model $C_2$}  \\
\colhead{} & \colhead{No Exchange}  & \colhead{With Exchange}  \\
}
\startdata 
Virial mass $M_{\rm vir}$ (\Msun) & 	$1.60\times 10^{12}$ & $1.43\times 10^{12}$\\
Virial radius $r_{\rm vir}$ (\kpc) & 	300 & 290\\
Halo concentration $C$ & 			12 & 12		\\
Disk mass $M_d$ (\Msun) & 		$7.0\times 10^{10}$ &$9.0\times 10^{10}$\\
Bulge mass $M_b$ (\Msun) & 		$1.9\times 10^{10}$ &$2.4\times 10^{10}$\\
Disk exponential scale $r_d$ (\kpc) & 	5.7 & 5.7 (8.55)\\
Stellar mass $r<3.5$~kpc (\Msun) & $2.8\times 10^{10}$& $3.6\times 10^{10}$\\
DM mass $r<3.5$~kpc (\Msun) & $2.5\times 10^{10}$ & $1.8\times 10^{10}$\\
Bulge $(M/L)_R$ (\Msun/\Lsun) & 3.0 & 3.8\\
Disk $(M/L)_R$ (\Msun/\Lsun) & 0.93 & 1.2\\
Maximum circular velocity $V_{c, {\rm max}}$ (\kms) & 269& 262\\
Max. halo circular velocity $V_{\rm halo, max}$ & 184 & 176\\
\qquad without compression (\kms) &		 \\
Mass inside 100~\kpc\ radius (\Msun) & 	$8.5\times 10^{11}$ &$8.1\times 10^{11}$\\
Spin parameter $\lambda$ &	0.036 & 0.057\\
Fraction of ``galactic'' baryons &0.56 & 0.80	 \\
\enddata 
\end{deluxetable} 

In Models $A_2$ and $B_2$ we used a different approach. Here we
started by assuming a disk mass of $M_3=6 \times 10^{10}\,\Msun$. This
mass is only 20-50\% more massive than the value used in our favored
models $A_1$ and $B_1$.  It is often quoted as the fiducial mass of
the disk \citep{bt87}.  We then tried to find an acceptable fit to the
data by changing parameters of the halo. Even assuming an extremely
low halo mass and unrealistically low concentration, both models
fail. In Model $A_2$, the maximum circular velocity and the surface
density at the solar radius are both too large.  Model $B_2$ is a bit
better, but still the local surface density is too large, the angular
momentum is uncomfortably large, and the mass inside 100~kpc is too
small. We can decrease the local surface density by reducing the
exponential length of the disk. But that has the side effect of
increasing the disk mass inside the central 5~kpc region. As a result,
the rotation velocity gets unacceptably large and the model fails
again.

Some cautions should be given regarding the bulge mass and the mass
distribution inside the central $\approx$3~kpc. The values $M_1$,
$M_2$ and $M_3$ are the masses of three inter-penetrating components,
which are the dominant components in the regions of the nucleus, the
bulge/bar and the disk respectively.  The $\rho_3$ component is a
constant-height thin exponential disk, which remains as such even
inside the central $\approx$3~kpc region occupied by the galactic
bar. This is unphysical and is just a simplification for the
convenience of the fitting. In reality the disk should be mixed with
the bar-bulge in the central region and should constitute a single
component with the distribution of the central non-axisymmetric bar.
These complexities are only important for a transitional region
between the bulge and the disk at radii 2--4~kpc, where the disk mass
is comparable to the bulge mass. Inside 2~kpc the disk is much smaller
than the bulge and it does not matter how we treat the disk.  

\begin{figure}[tb!]
\epsscale{1.0} \plotone{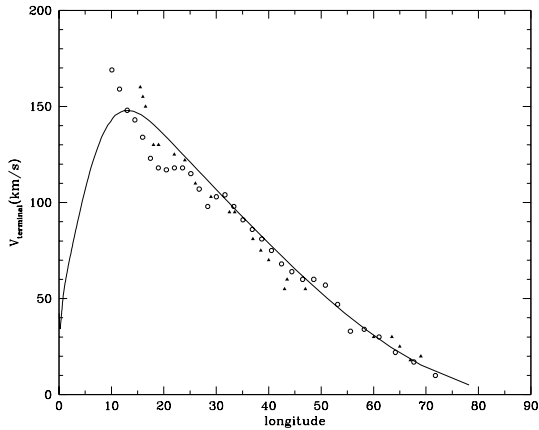}
\caption{\small Dependence of terminal velocities on galactic
longitude $l$. The full curve is for model $A_1$. Symbols show
observational data from the HI measurements of Knapp et al. (1985)
(circles) and Kerr et al.(1986) (triangles). At small angles the
deviations from circular velocities are expected to be large due to
the central bar. This is clearly seen at $l <
20^o$.}\label{fig:TermVModelA}
\end{figure}

\begin{figure}[tb!]
\epsscale{2.2}
\plottwo{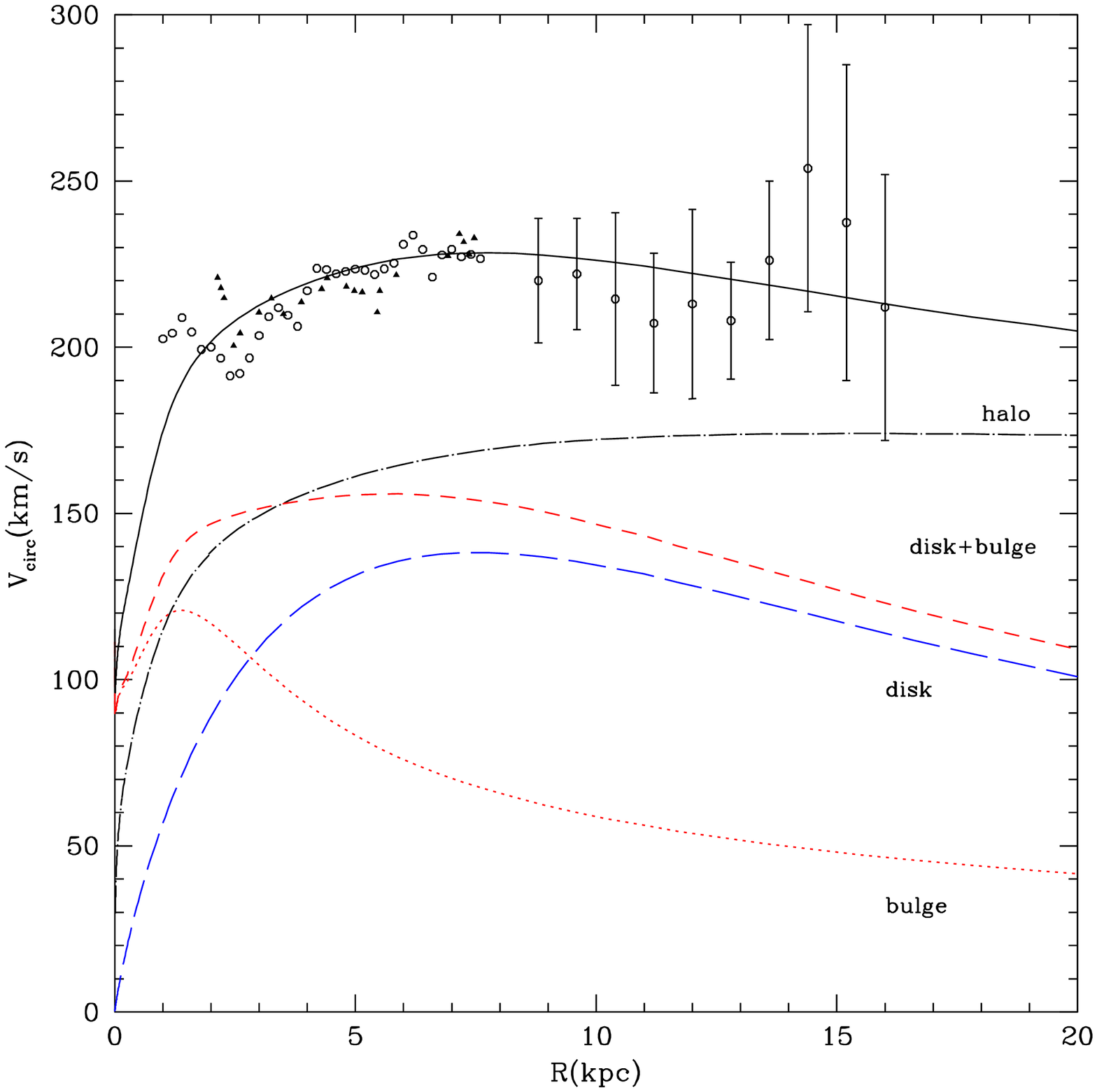}{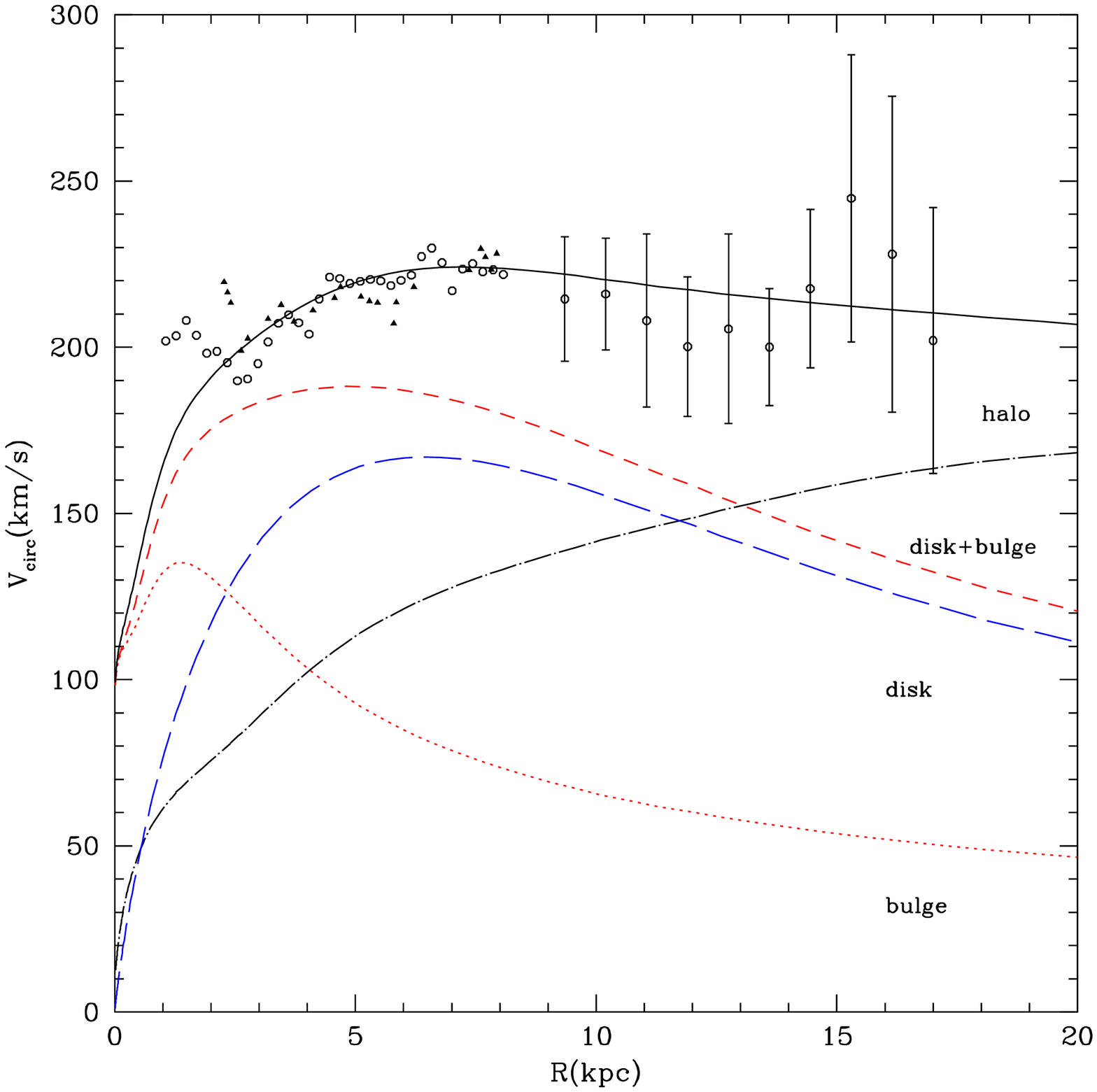}
\caption{\small Rotation curve for our favorite models $A_1$ (no exchange
of angular momentum) and $B_1$ (with the exchange). Note that the dark
matter dominates only in the outer part of the Milky Way. Symbols show
observational data from HI measurements of Knapp et al. (1985)
(circles) and Kerr et al.(1986) (triangles). 
}\label{fig:RotModelA}
\end{figure}

\begin{figure}[tb!]
\epsscale{1.}  \plotone{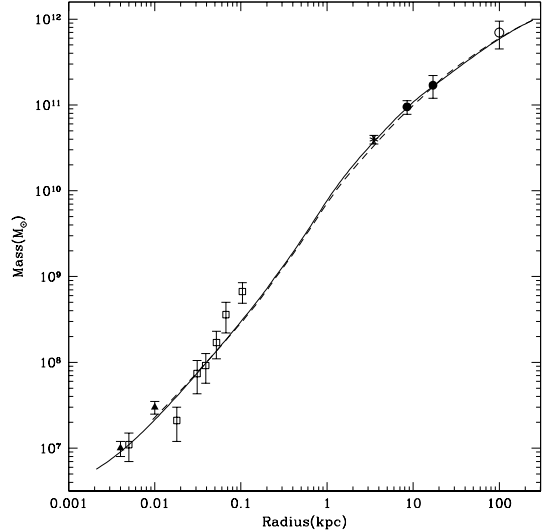}
\caption{\small Mass distribution of the MW galaxy for Model $A_1$ (full
curve) and model $B_1$ (dashed curve). The large dots with error bars are
observational constraints. From small to large radii the constraints
are based on: stellar radial velocities and proper motions in the
galactic center; radial velocities of OH/IR stars; modeling of the bar
using DIRBE and stellar velocities; rotational velocity at the solar
radius; dynamics of satellites.  }\label{fig:massMW}
\end{figure}

\begin{figure}[tb!]
\epsscale{1.05} \plotone{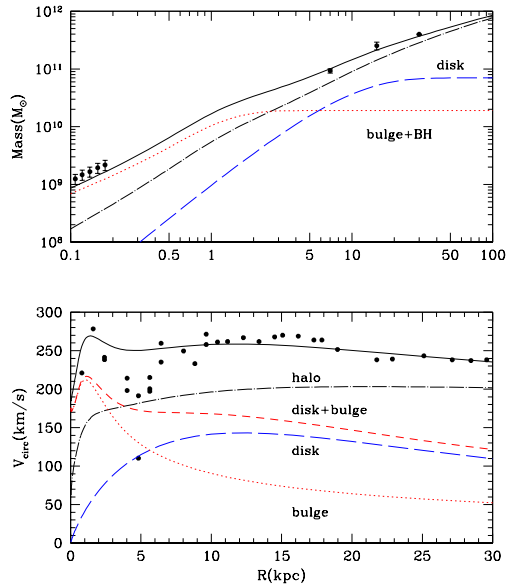} 
\caption{\small Mass distribution ({\it top panel}) and rotation
velocity ({\it bottom panel}) of the M31 galaxy for Model $C_1$. The
large dots with error bars are observational constraints.  In the
bottom panel the circles show results of CO ($r< 10~\kpc$) and HI ($r>
10~\kpc$) observations. The large dip at around 5~kpc is likely due to
non-circular motions induced by the bar. Observational data points in
the top panel at small radii are from stellar motions in the
nucleus. Vertical bars correspond to 20\% errors in mass. Data points
at 7, 15, and 30~kpc correspond to circular velocities of $240\pm 10,
270\pm 20,$ and $240\pm 10$\kms.  }\label{fig:RotationM31}
\end{figure}

\begin{figure}[tb!]
\epsscale{1.1}
\plotone{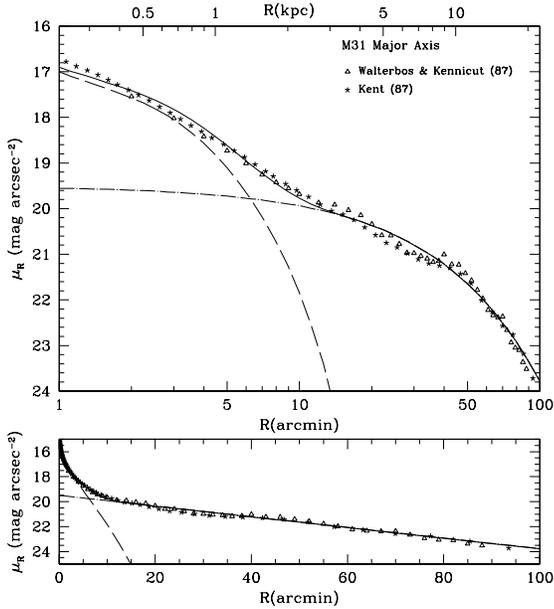}
 \caption{\small Surface brightness of M31 in the R-band on linear ({\it
 bottom panel}) and logarithmic ({\it top panel}) scales. Deviations
 from the observational results are less than 0.2~mag. 
 }\label{fig:brightnessM31}
\end{figure}

Some important quantities for the best models of the Milky Way and M31
are presented in Figures \ref{fig:TermVModelA} --
\ref{fig:brightnessM31}.  Figure~\ref{fig:TermVModelA} compares the
observed terminal velocities in our Galaxy with the predictions of
model $A_1$. The quality of the fit decreases at small angles because
of non-circular motions produced by the central bar. Even larger
distortions are found in the rotation curve of M31, which has a large
bar. The usual way to deal with the deviations is to ignore the data
at small angles. For example, \citet{DehnenBinney98} ignore data for
angles smaller than 17$^o$, which corresponds to $\approx
2.5$~kpc. \citet{OllingMerrifield98} used the data of
\citet{Malhotra95}, which also start at $\approx 2.5$~kpc. Without a
more realistic (non-axisymmetric) model of the bar one cannot do
better than this. The problem is that we can not exclude the
possibility that small $5-10~\kms$ disturbances related to the bar can
be felt even at larger radii. This could bias our parameter estimates,
which are sensitive to what happens at those radii. For this reason,
we still consider the fit produced by model $A_1$ to be acceptable in
spite of the fact that for longitudes $20^\circ-30^\circ$ the model
values exceed the observational points by $5-10~\kms$.

Figure \ref{fig:RotModelA} shows the rotation curve of our Galaxy. The
observational data are compared with models $A_1$ and $B_1$. We also
show the contributions of different components. For both models the
dark matter dominates in the outer parts of the Galaxy, but the radius
at which the contribution of the dark matter equals that of the
baryonic component are very different. For model $A_1$, the masses
contributed by baryons and dark matter are equal at 3.5~kpc. For model
$B_1$ this radius is 11~kpc. The difference is due to a combination of
two factors: expulsion of dark matter by dynamical friction and an
increase in the assumed disk+bulge mass. In the central 5~kpc of the
Galaxy the contribution of the dark matter relative to the disk and
bulge is very different for models $A_1$ and $B_1$. In the case of
model $A_1$ and in all other models with no exchange of angular
momentum, the dark matter is never a strongly dominant component, but
it always contributes 40\%--60\% of the total mass. In other words,
these models have sub-maximal disks \citep{Bottema97, CourteauRix}.
In the case of models with exchange of the angular momentum, there is
little dark matter in the central region of our Galaxy.

Figure \ref{fig:massMW} shows the distribution of mass in the Milky
Way galaxy for a very large range of scales, from 5~pc to 200~kpc. The
following observational constraints are used. The first two points at
5-10~pc are from studies of stellar radial velocities and proper
motions in the galactic center. Squares are based on kinematics of
OH/IR stars \citep{Lindqvist1992}. The point at 3.5~kpc is based on
the \citet{Zhao96} model of the bar. Because the model was compared
with the data on stellar kinematics (inner rotation curve and radial
velocity dispersion), it gives a constraint on the total mass:
$4\times 10^{10}\, \Msun$, with an uncertainty of about 20\%.  For the
next data point at 8.5~kpc we simply assume that the circular velocity
is $220\pm 20~\kms$, which covers the whole range of reasonable
values. We then estimate the mass as $M=v^2r/G$. The last
observational point is the constraint from the motions of satellite
galaxies discussed in section \ref{sec:ObsConstraints}. The central
data points were not used in either our fitting or in the analysis of
the bulge \citep{Zhao96}. Nevertheless, they come fairly close 
to the extrapolation of our model into the very
center of our Galaxy.  The theoretical curves for our favored models $A_1$
and $B_1$ are very close to each other, which is not surprising because
they fit the same data and have the same global dark matter
content. The largest deviation of the models from the data is for the
mass inside 100~pc, where the observational estimate is twice larger
than the prediction of the models.  Even at this point the
disagreement is not alarming because the observational data are likely
more uncertain than the formal error.

What is remarkable about Figure \ref{fig:massMW} is that it spans more
than 5 orders of magnitude in radius and mass. It is encouraging that,
without fine-tuning, our models are consistent with observations of
the dynamical mass of the MW over this huge range.

Finding an acceptable model for M31 was relatively easy because there
are much less data. In particular, we do not have kinematic
constraints for the disk, which would be equivalent to constraints at
the solar position in our Galaxy. Our model seems to reproduce
reasonably well the dynamical mass of M31 from 100~pc to
$\approx$100~kpc. Our model does not produce the very large wiggles
exhibited by the observed rotation curve. The wiggles at 5~kpc and
9~kpc are likely due to non-circular motions induced by the bar and,
thus, as discussed before, cannot be reproduced by any axisymmetric
model.  The bulge of M31 is almost twice as massive as the bulge of
our Galaxy. It is also slightly (30\%) more compact. The disk of M31
is also more massive, but it is more extended. As a result, in the
central 5~kpc of the M31 the bulge is a much more dominant component
as compared with the bulge of our Galaxy.

The surface brightness profile in the R-band, shown in
figure~\ref{fig:brightnessM31}, is used as an additional constraint.
An accurate fit (the same as for the mass modeling) is obtained for
stellar mass-to-light ratios of $M/L=0.93\,\Msun/\Lsun$ and
$M/L=3\,\Msun/\Lsun$ for the disk and the bulge respectively.  These
results are quite consistent with the expectations from stellar
population synthesis models for a galaxy with the $B-R$ color of M31,
which imply an overall stellar mass-to-light ratio $(M/L)_R =
0.85\,\Msun/\Lsun$ \citep{bell-dejong}.  The bulge $M/L$ is likely
overestimated, and would be reduced by a correction for internal dust
absorption. For example, internal absorption of 0.25~mag inside the
bulge would reduce $M/L$ to 2.4. At a first glance our estimates of
the M/L ratios are significantly different from those of
\citet{Kent89}, who give $M/L=5\,\Msun/\Lsun$ and
$M/L=10.4\,\Msun/\Lsun$ for the bulge and the disk. Most of the
differences come from differences in bands ($r$ instead of R) and from
the fact that Kent's luminosities were not corrected for
absorption. If we include those corrections (see
Section~\ref{sec:obsM31}) and also scale the results to the distance
of 770~pc, Kent's estimates become $M/L=2.5\,\Msun/\Lsun$ for the disk
($M/L\approx 6.7$ in B) and $M/L=2.4\,\Msun/\Lsun$ for the bulge. Such
a large M/L ratio for the disk, more typical of an old bulge
population, seems a bit problematic. The main reason why
\citet{Kent89} has such a large M/L lies in the assumed profile of the
dark matter: constant density through the whole galaxy. This results
in the extreme case of a maximal disk, and very little dark matter
inside 20-25~kpc.

\subsection{Effects of compression by baryonic infall}
\label{sec:compression}
The sinking of baryons into the central part of a galaxy increases the
depth of the gravitational potential. This leads to an increase of the
dark matter density. To some degree, this process provides coupling of
the baryons and the dark matter. As a result, in the case of adiabatic
compression without the exchange of the angular momentum, it appears
difficult to find a case where one of the components (dark matter or
baryons) is significantly larger than the other. Figure~
\ref{fig:RotModelA}, top panel, gives an example of this effect. From
100~pc to about 6~kpc the contributions of baryons and dark matter to
the circular velocity are about equal.

\begin{figure}[tb!]
\epsscale{1.23}
\plotone{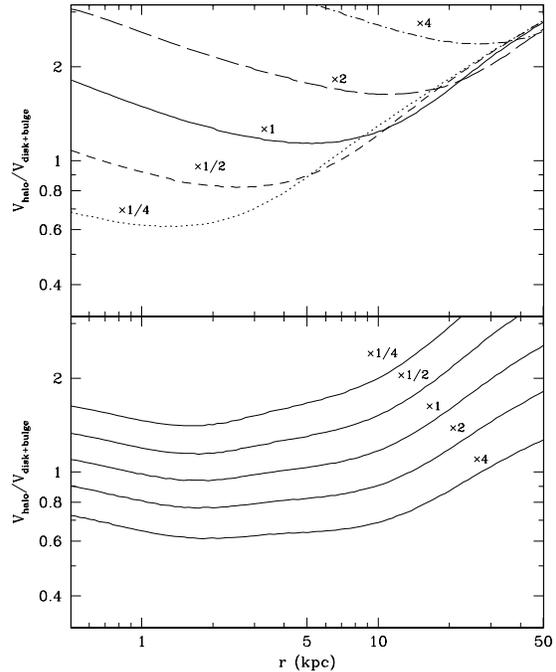}
\caption{\small The ratio of the circular velocity produced by the
dark matter to that of the baryons for two models of adiabatic
compression. For the {\it top panel} the exponential disk scale length
$r_d$ was changed by a factor indicated in the plot. The factor of
unity ($\times 1$) is for the assumption of a disk mass $4\times
10^{10}\, \Msun$ and scale length $r_d=3.5$~kpc. This model has no bulge.
Curves in the {\it bottom panel} show the same quantity when the disk
scale length is held fixed at $r_d=3.5$~kpc the total baryonic mass is
varied. Parameters of the $\times 1$ curve are the same as for Model
$A_1$.  }\label{fig:Compression}
\end{figure}

\begin{figure}[tb!]
\epsscale{2.0}
\plottwo{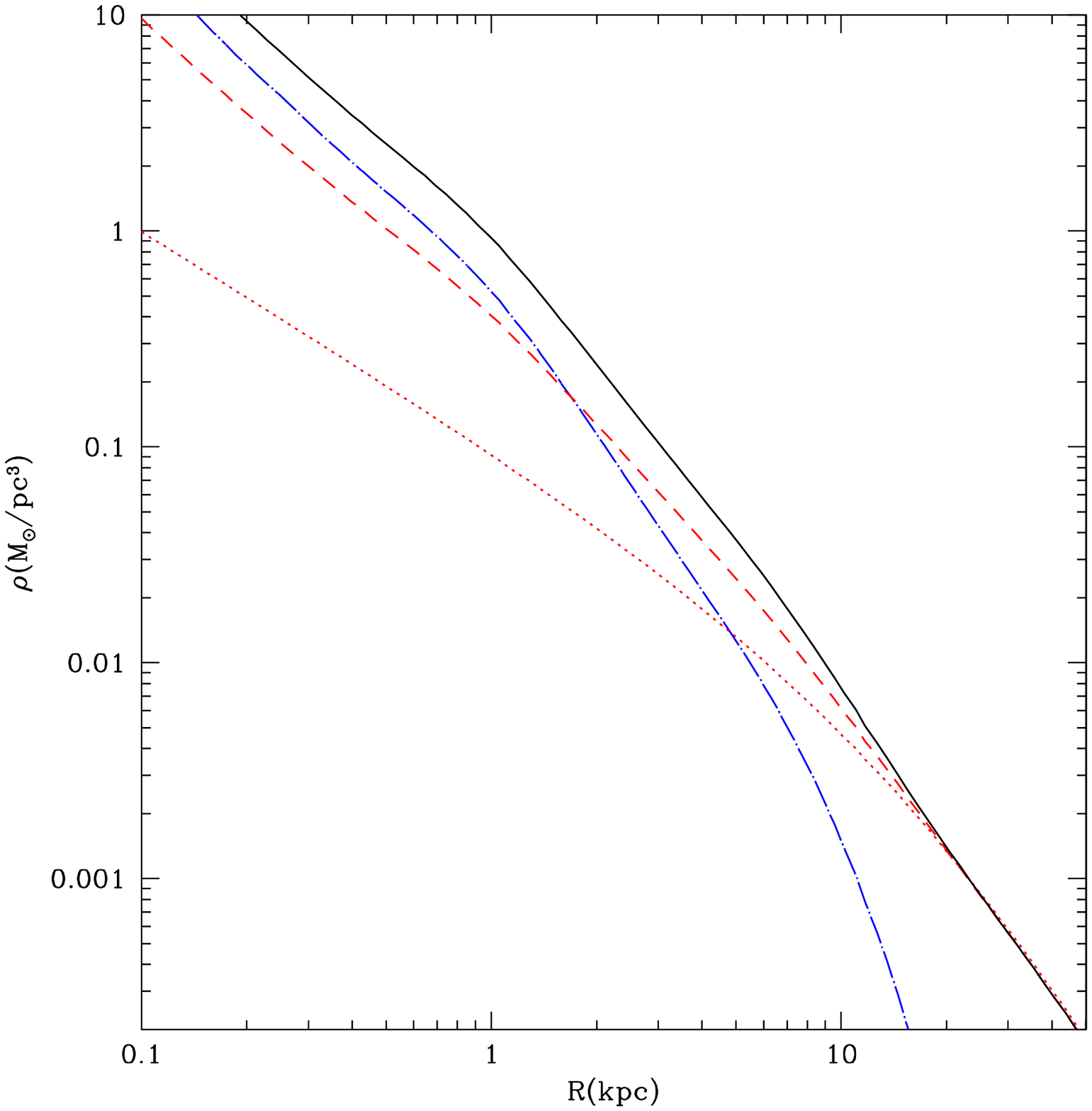}{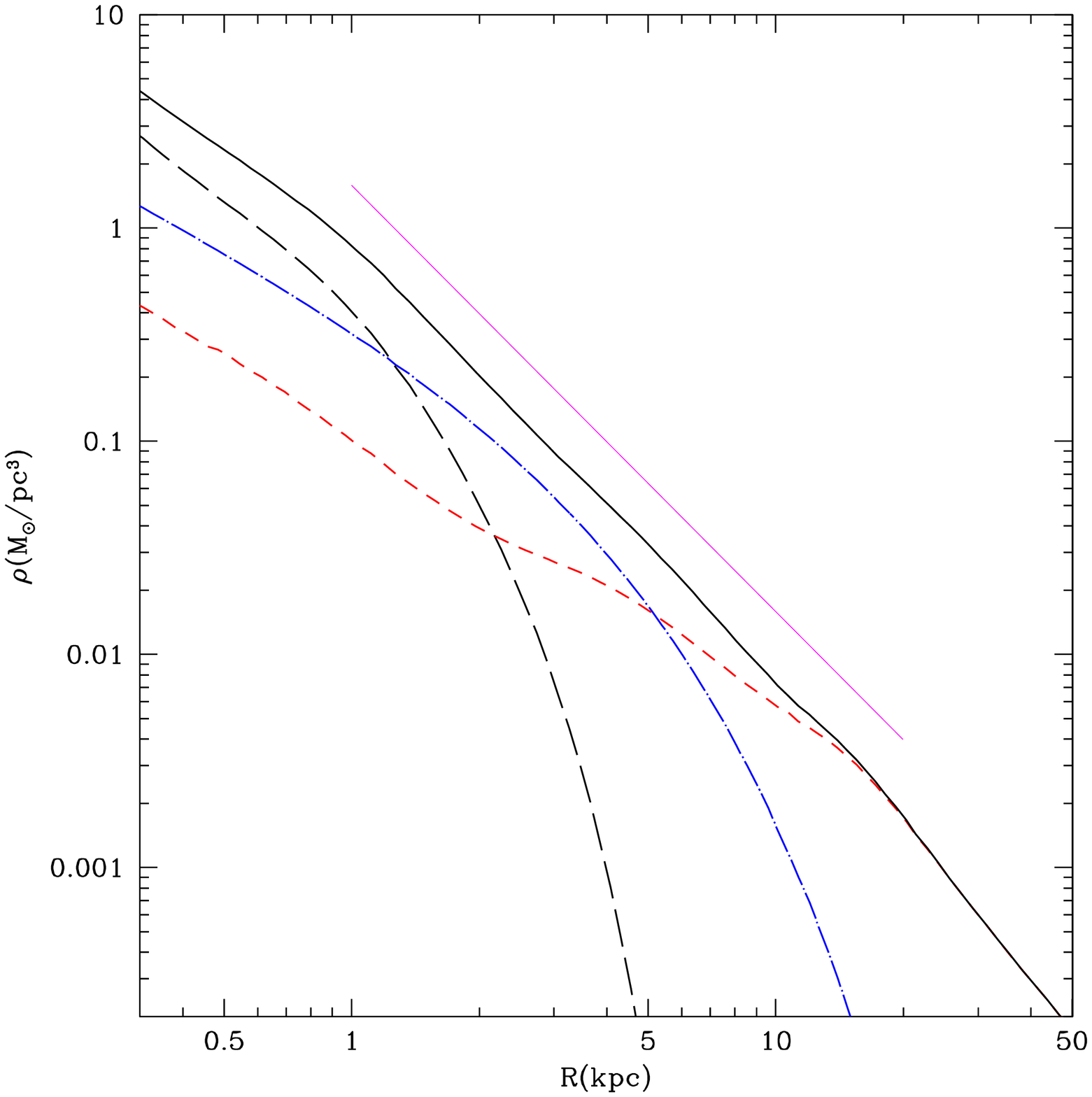}
\caption{\small Density distribution for different components in
models $A_1$ (top panel ) and $B_1$ (bottom panel. Different curves
show: the total density (full) and dark matter (short dash). {\it
Top:} The total bulge+disk density is shown by the dot-dashed
curve. Inside $\approx 3$~kpc the baryons and the dark matter closely
follow one another. This illustrates the self-adjusting nature of the
compression. The dotted curve shows the dark matter before the
compression by baryons. The compression is clearly a dominant effect
inside the central 10~kpc. {\it Bottom:} The bulge is shown by the
long-dashed curve; the disk is shown by dot-dashed curve. The
components have very different profiles, but their sum is very close
to a power-law with slope -2, shown by the straight line. This
``disk-halo conspiracy'' is the natural result of the adiabatic dark
matter compression.}\label{fig:Density}
\end{figure}

We use two simple models to study the effect of adiabatic compression
on the ratio of dark matter and baryons. In the first case we take
Model $A_1$ and change the mass of baryons while keeping the disk scale
length and the bulge-to-disk ratio constant. In the second case, we
vary the scale length of the disk, keeping the mass of the disk the
same as in Model $A_1$; we ignore the bulge to make the case more clear.
Figure \ref{fig:Compression} shows the ratio of the circular
velocity produced by the dark matter to that of the baryons.

Figure \ref{fig:Compression} illustrates that for a realistic range of
parameters the contribution of baryons can not be significantly larger
that of the dark matter. Even in the case when disk is four times more
massive or four times more compact than in Model $A_1$, the contribution
of the dark matter to the rotation curve was at least 1/3.  In other
words, in hierarchical models with adiabatic compression and no
exchange between components it is difficult to have a maximum disk.

It is interesting to note that the opposite is also true: in the
central part of a galaxy it is difficult to have a model where the
mass of the dark matter is much larger than that of baryons. The model
with a four times more extended disk in the top panel ($\times 4$) is
an example of what should represent a low surface brightness
galaxy. Its surface density is 16 times smaller than for our Galaxy,
and its mass in stars is only 1\% of the virial mass, yet the baryons
at the solar radius contribute about 1/4 of the rotational velocity.

In view of this self-adjusting behavior of the dark matter, it is
easier to understand the profiles of different components in realistic
models.  The top panel in Figure~\ref{fig:Density} shows the dark
matter and the combined disk and bulge density profiles. For
comparison we also show the dark matter profile before the baryonic
infall. Note that the change in the dark matter profile is quite
large, yet the final profile manages to follow the baryonic density
very closely in the central region. The bottom panel shows different
components for model $B_1$. Each component has a complicated
shape. Different components become prominent at different radii. The
net effect is rather remarkable --- the sum of all components is very
close to a power-law with slope -2. The deviations from this power law
are less than 10\% on scales of 1--20~kpc. One may argue that this
occurs because we are fitting a rotation curve that is nearly flat,
and so we get a $r^{-2}$ density profile. This is not exactly
true. The observed deviations are quite substantial. The final fit is
much closer to a pure power law than what is required by the
observations. This ``disk-halo conspiracy'' is the natural result of
the coupling of baryons and dark matter via the adiabatic compression.

\subsection{Constraints from the Global Luminosity Function}
\label{sec:lf}
If the luminosity of our Galaxy is typical of that of other disk-type
galaxies that form in halos of the same mass, this also has
implications for the global luminosity function. We now investigate
the number density of galaxies like the Milky Way, using the revised
Press-Schechter approximation of \citet{shethtormen:99}. The
Sheth-Tormen approximation provides us with the number density of dark
matter halos as a function of their mass. Once we have pinned down the
luminosity of the Milky Way, we can calculate the mass-to-light ratio
within the virial radius of the dark matter halo hosting the galaxy in
a given model. We can then convert the halo number density per mass
interval to a galaxy number density per magnitude interval, which may
be compared with the observed luminosity function. We neglect the
contribution from galaxies that are ``satellites'' in much larger mass
halos (i.e., cluster galaxies), and assume that the adopted
mass-to-light ratio holds for disk galaxies only. We then compare this
prediction with the K-band luminosity function for all galaxies and
for late-type galaxies from the 2MASS survey \citep{lf2mass}.

We use two different approaches to estimate the K-band luminosity of
the Galaxy. The total luminosity of our Galaxy is notoriously
difficult to measure directly because we live in the middle of it, and
obscuration is important even in the K-band. \citet{DrimmelSpergel2001}
recently performed a detailed analysis in which they estimated the
K-band luminosity of the Milky Way by fitting a model to the COBE
DIRBE photometric data. They found $M_K = -24.0$ for the total
luminosity of the Galaxy.

The other approach is to assume that the Galaxy is ``typical'' and
should lie on the Tully-Fisher relation (TFR). Unfortunately, this
approach is also uncertain because there is little Tully-Fisher data
in the K-band and the results in the literature are somewhat
inconsistent. If we use the TFR of \citet{malhotra96}, assuming that
the velocity width of the Milky Way $\Delta V = 2V_c = 440$ km/s, then
we obtain $M_K = -23.37$, about 0.7 magnitudes fainter than the direct
DIRBE measurement. If we use instead $\Delta V = 480$ km/s, which
Malhotra et al. obtain by scaling from M31 ($\Delta V (\rm{MW}) =
(220/250) \Delta V({\rm M31})$, with $\log_{10}(\Delta V({\rm
M31}))=2.737$), then we get $-23.69$, only 0.3 magnitudes fainter than
the direct measurement. This would place the MW about 1-$\sigma$
brightwards of the TFR, which seems not unreasonable. However, the
K'-band TFR recently published by \citet{tp00} and \citet{rothberg:00}
gives $M_K = -24.4$ for $\Delta V = 440$ km/s. The slope is very
similar to the relation of Malhotra et al., but the zero-point is
almost one full magnitude brighter. This problem is only exacerbated
if we consider that the Cepheid calibration data used by Malhotra et
al. is consistent with a Hubble parameter of $H_0 = 71$ km/s/Mpc,
while the ground-based $K'$ data imply $H_0 = 81$ km/s/Mpc. If we
scale the zero-point of the ground-based data to the lower value of
$H_0 \simeq 70$ km/s/Mpc, all the galaxies are further away and
therefore are even brighter, making the discrepancy worse (the Milky
Way would then have $M_K = -24.7$).

\begin{figure}[tb!]
\epsscale{1.04}
\plotone{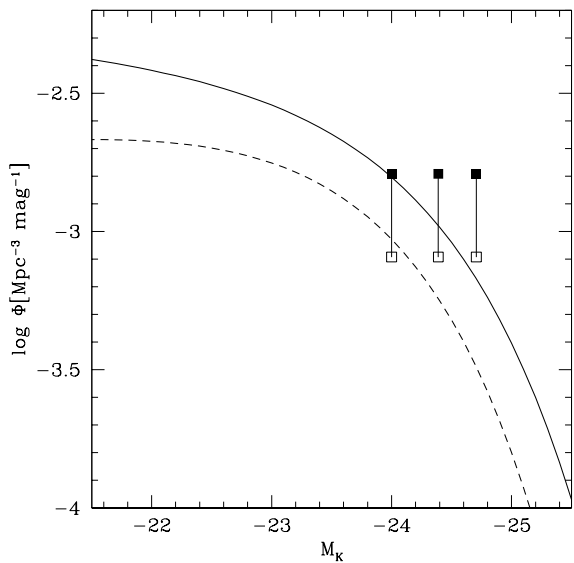}
\caption{\small The K-band galaxy luminosity function. The curved
lines show the observed K-band luminosity function from the 2MASS
survey (Kochanek et al. 2001) scaled to $H_0 = 70$ km/s/Mpc; the solid
line shows the results for all types of galaxies and the dashed line
shows the results for late-type galaxies only. Points show the
prediction of the number density of halos hosting ``Milky Way''
galaxies, using the halo virial mass from model $A_1$ or $B_1$ and an
assumed Milky Way luminosity. Solid symbols show the value obtained if
all halos are assumed to host disk galaxies, and are to be compared
with the solid lines. Open symbols show the result obtained if only
half of the halos host disk galaxies, and are to be compared with the
dashed lines. The three sets of connected points are for three
different assumed values for the total K-band luminosity of the Milky
Way (see text). }
\label{fig:lfnorm}
\end{figure}

\citet{malhotra96} base their TFR on photometry from DIRBE and have a
total of seven galaxies in their sample (including the Milky Way). The
ground-based K'-band sample considered by \citet{rothberg:00} and
\citet{tp00} are from the Ursa Major sample \citep[see
also][]{verheijen:01} and additional galaxies from the Pisces
filament, and consists of a total of 69 galaxies with 4 galaxies
having Cepheid distances. All of these galaxies have multi-band
photometry in B, R, I and K' and the TF relations in the optical bands
agree well with those from other samples, including the very
well-studied I-band. Therefore it seems that the zero-point for the
ground-based sample should be more reliable. This suggests that
perhaps there is some sort of calibration offset in the DIRBE
photometry, and perhaps calls into question the validity of using even
the \citet{DrimmelSpergel2001} result. We shall consider the
implications of using three different choices for the total K-band
luminosity of the Galaxy: the direct DIRBE result, the raw
ground-based TFR result, and the ground-based TFR result scaled to
$H_0 = 70$ km/s/Mpc. The three sets of squares plotted in
Figure~\ref{fig:lfnorm} correspond to these three values.

In each case the higher of the two connected squares shows the number
density that we would obtain if every halo hosted a disk galaxy. The
lower point shows the value obtained if only half of the halos host
disk-type galaxies (probably a reasonable lower limit, based on
semi-analytic modelling).  The points can easily be scaled up or down
for different assumptions about the fraction of disk-hosting halos,
but for reasonable assumptions the result should lie between the lower
and upper points. Unfortunately, the interpretation of the luminosity
function constraint is quite sensitive to the uncertain value of the
MW's total luminosity. For the bright normalization (from the TFR of
\citet{rothberg:00} scaled to $H_0 = 70$ km/s/Mpc), the number density
of halos harboring MW galaxies is about a factor of two higher than
the observations. For the faint normalization \citep[the direct value
from][]{DrimmelSpergel2001}, the number density of MW halos is
marginally consistent with the observed luminosity function.

\begin{figure}[tb!]
\epsscale{1.04}
\plotone{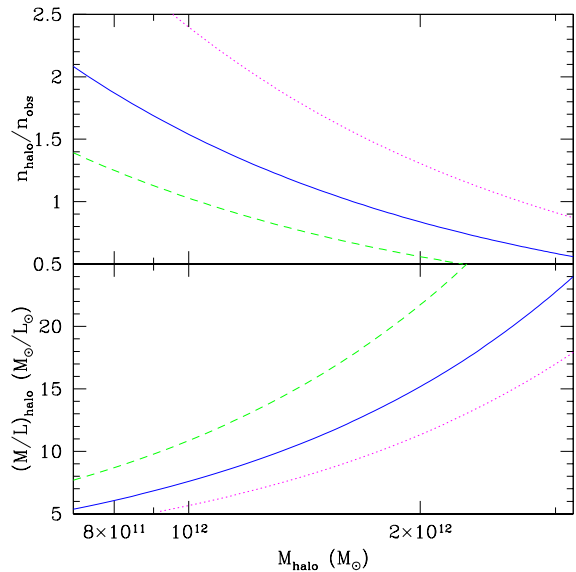}
\caption{\small Constraints on the virial mass of the Milky Way from
the observed luminosity function. The top panel shows the ratio of the
predicted to observed number densities of Milky Way galaxies as a
function of the assumed average virial mass of dark matter halos
hosting galaxies with the luminosity of the Milky Way. The three lines
correspond to the three values of the Milky Way luminosity discussed
in the text; where the dotted line is the bright normalization, the
solid line is the intermediate normalization and the dashed line is
the faint normalization. The plotted line should be less than unity in
order for the model to satisfy the luminosity function
constraint. Larger masses have lower number densities and therefore
more easily satisfy the constraint.  The bottom panel shows the halo
mass-to-light ratio (i.e. the halo virial mass divided by the total
galaxy luminosity) as a function of halo mass. Again, the three lines
correspond to the three different Milky Way luminosities considered in
the text.}
\label{fig:mwmass}
\end{figure}

It is of some concern that for the Milky Way normalization that we
regard as the most reliable (the ground-based TFR scaled to the
standard value of $H_0$), the luminosity function constraint is
violated. If this normalization turns out to be correct, then this may
suggest that the Milky Way is harbored by a halo with a larger virial
mass. The number density of dark matter halos drops fairly rapidly
with virial mass so this relaxes the constraint
somewhat. Figure~\ref{fig:mwmass} shows the number density of dark
matter halos as a function of their virial mass, normalized by the
observed number density of galaxies at the assumed luminosity of the
Milky Way from \citet{lf2mass}. We see that the faint and even the
intermediate Milky Way luminosity can be accomodated fairly
comfortably within the mass range allowed by our dynamical modeling
($10^{12}\, \msun < \Mvir < 2\times10^{12} \,\msun$; with brighter Milky
Ways favoring larger virial masses), however, the bright Milky Way
normalization is too high even at the upper virial mass limit of
$2\times10^{12}\, \Msun$.

\section{Discussion}
\label{sec:discussion}
There are numerous aspects of these models of the MW and the M31
galaxies which are important to consider. We start with very small
scales and proceed to larger ones.

\subsection{Central region and the black hole}
It is now quite well established that black holes of mass $\sim
10^6-10^8\, \msun$ reside in the centers of most galaxies.  The case for
a $2.6\times 10^6\, \msun$ black hole at the center of the Milky Way is
demanded by rising radial velocity and proper motion dispersions
within a few pc of the center (\citet{Ghez00} and references therein).
Recently, \citet{GS99} showed that if the BH is grown adiabatically,
the central dark matter will be drawn into a dense spike, with density
exceeding $10^8\, \Msun/\pc^3$.  This would make the Galactic center very
luminous in terms of neutrino flux if the dark matter is made of
annihilating neutralinos.  \citet{UZK01} however argue that it is
unlikely for the BH to grow all its mass while at rest in the center
because of galaxy merging in CDM models.  A spiraling-in BH would
actually ``heat'' and reduce the central density of the dark matter.
The seed is likely to be massive ($\ge 10^5\, \Msun$), because otherwise
it would not have enough time to spiral into the center by dynamical
friction.  Realistic models of loss-cone capture show that the rate of
stellar capture is around $10^6\, \Msun$ per Hubble time
\citep{MagorrianTremaine}. This implies that the BH could at most
double its mass during the adiabatic phase.

Following \citet{UZK01} we take into account the effect of adiabatic
growth from a seed BH of half the final value. The flux is very likely
to be small because in our models the final peak of the dark matter
density outside the event horizon of the BH is well below $10^8\,
\Msun/\pc^3$, which is the amount needed for significant annihilation.

\subsection{Microlensing Constraints}
\label{sec:microlensing}
Microlensing events toward the Galactic center provide a lower limit
on the mass of baryons within the inner part of the Galaxy. Combined
with the observed terminal velocities, this places upper limits on the
mass of dark matter that can be present and implies that the inner
Galaxy must be mostly baryonic with a massive bulge and disk.  The
microlensing events toward the Galactic bulge are still not completely
understood observationally and theoretically, in particular the nature
of two long-duration events near the $l=3^\circ$, $b=-3^\circ$ field.
Several values for the spatially averaged optical depth are reported
with large discrepancies in the mean values.  The earlier values are
too high for axisymmetric bulge and disk models, and are barely
consistent even with a massive bar.  A $2.2\times 10^{10}\, \Msun$ bar
pointing about $13^\circ-20^\circ$ away from the Galactic center-Sun
line plus a full disk with a mass of $1.7\times 10^{10}\, \Msun$
inside 3.5 kpc are required by detailed dynamical models of the COBE
map and by the kinematics of the bulge stars
\citep{Zhao96}. \citet{ZSR95} found that the bulk of the lenses are on
the near side of this elongated bar.  This bar plus disk model can
produce an optical depth consistent with the lower $1\sigma$ error bar
of the observed microlensing optical depth of $(3.3 \pm 1.2)\times
10^{-6}$ for all stars observed by OGLE \citep{udalski:94} toward
Baade's window ($l=1^\circ$ and $b=-4^\circ$). \citet{ZRS96} found
that this model with a flat IMF with
very few brown dwarfs can fit the microlensing event duration
distribution.  \citet{Peale} studied the optical depth and the
distribution of durations of lensing events for the bulge with the
shape used in our paper. 
Both statistics were found to be compatible with observational data.

\cite{ZhaoMao96} pointed out that except for very contrived models,
the higher reported optical depth values are in conflict with the
total lensable material in the inner Galaxy.  The optical depth for
the low-latitude red clump subsample from MACHO \citet{alcock:96} is
between $(3-10) \times 10^{-6}$ at the 95\% confidence level.  This would
require contrived models involving the Sun-center line being the
diagonal line of a homogeneous rectangular bar.  This discrepancy has
been highlighted by \citet{kuijken} and by \citet{BBG00} for
axisymmetric models.  Both confirmed that the need of extremely
contrived bar and disc models if the optical depth is as large as
$(3-10) \times 10^{-6}$.

The most recent analysis of bulge microlensing, however, suggests a
lower optical depth in general.  Statistics have improved as we now
have over 500 events.  In particular, the value for red clump giants
is the range $(1.4-2) \times 10^{-6}$ depending on whether the long
duration events are included (Popowski 2001, private communications).
\citet{SevensterKalnajs:01} suggested that the 3 kpc arm might
correspond to a stellar ring, and also account for a large fraction of
the microlenses.

Without making detailed models of the non-axisymmetric bar and the
ring, we can do a quick check of whether our models are consistent
with the microlensing data by scaling the results of \citet{ZRS96} and
\citet{Peale} for the masses of baryonic components in our models.  If
we use only the $\rho_2$ bar component (mass $10^{10}\, \msun$), then
the optical depth is $\sim 10^{-6} (M_2/10^{10}\, \Msun) \sim 10^{-6}$
\citep{ZSR95}.  In our model $A_1$ there is another $\sim 10^{10}\,
\Msun$ of baryons inside 3.5~kpc which is formally in the $\rho_3$
axisymmetric disk. As we have discussed, this thin unperturbed disk
inside the non-axisymmetric bar is only a mathematical
simplification. It is unlikely that it can stay thin or even can
survive inside the bar. The most probable scenario is that the bar was
formed from the disk at some epoch, when the disk became unstable. In
this case the disk becomes the bar and we should expect that the inner
disk has the same distribution as the bar. In this case, the
microlensing counts should include all the baryonic mass inside
3.5~kpc, which doubles the estimate of the optical depth.  Then our
best models $A_1$ and $B_1$ are consistent with an optical depth of
$(1.4-2)\times 10^{-6}$.

\subsection{Comparison with other work} 

Our estimate for the virial mass of the M31 halo $1.4
\times10^{12}\, \Msun$ is consistent with earlier estimates which used
kinematics of satellites.  \citet{CourteauBerg99} found a total mass
of $(1.33\pm 0.18)\times 10^{12}\, \Msun$ inside $260~\kpc$, and
\citet{EvansetalM31} and \citet{EvansWilkinson2000} found
$(1.23^{+1.8}_{-0.6})\times 10^{12}\, \Msun$.  The main difference
between our results and \citet{EvansetalM31} is that they assumed a
halo density profile very different from an NFW profile.

For the Milky Way halo, our models $A_1$ and $B_1$ predict a mass
$M_{R<100\kpc}=(5.8-6) \times 10^{11}\, \Msun$, consistent with what
\citet{DehnenBinney98} find in their ``standard models 1-4'',
$M_{R<100\kpc}= (6.0-6.6)\times 10^{11}\, \Msun$.  Our disk also has a
mass of $(4-5)\times 10^{10}\, \Msun$, consistent with their finding
of $M_d= 4.2- 5.1\times 10^{10}\, \Msun$.

Our results and conclusions differ from those of
\citet{Hernandez2001}, who model the Milky Way in the framework of a
cosmological scenario. \citet{Hernandez2001} find that the standard
halo profiles are not compatible with the observed properties of the
Milky Way and that ``the rotation curve for our Galaxy implies the
presence of a constant density core in its dark matter halo''. The
gross features of our approach and that of \citet{Hernandez2001} are
similar. The parameters of the cosmological models are only slightly
different. Adiabatic infall corrections are also used, and the dark
matter profiles are also similar. The main difference was in the mass
of the halo assumed for the Galaxy. The average virial mass in the
models of \citet{Hernandez2001} was assumed to be $M_{\rm vir}
=2.8\times 10^{12}\, \Msun$, which is almost three times larger than the
virial mass in our favored models $A_1$ and $B_1$. If we ran our models for
this large virial mass, we would also reject the model (our model $A_4$
has the largest virial mass, $M_{\rm vir} =2.0\times 10^{12}\, \Msun$,
and even this model is marginal, with a rather large rotation velocity
and an abnormally low concentration).  The reason that
\citet{Hernandez2001} adopted such a large virial mass is their
perhaps overly stringent interpretation of the \citet{Kochanek96}
estimate of the mass inside 50~kpc. \citet{Hernandez2001} assume that
the mass was $M_{50} =(4.9\pm 0.5)\times 10^{11}\, \Msun$, while
\citet{Kochanek96} gives $M_{50} =(3.2-5.5)\times 10^{11}\, \Msun$ at the
90\% confidence level if Leo I is excluded from the analysis. For
comparison, our models $A_1$ and $B_1$ have $M_{50} =(3.7-3.8)\times
10^{11}\, \Msun$, which is compatible with
\citet{Kochanek96}. \citet{WilkinsonEvans} give even larger
uncertainties on the mass: $M_{50} =5.4^{+0.2}_{-3.6}\times
10^{11}\, \Msun$.

There is no doubt that estimates of the mass at large radii, 50~kpc --
100~kpc, are crucial for constraining the halo parameters. Values of
$M_{50}$ larger than $5\times 10^{11}\, \Msun$ are very likely to be
difficult to reconcile with the standard cosmological models. The
problem is that the observational constraints are still very much
uncertain.

It is interesting to compare our results with the results of
\citet{BE2001}, who came to the conclusion that ``cuspy haloes favored
by the Cold Dark Matter cosmology (and its variants) are inconsistent
with the observational data''. Their main argument was a combination
of three constraints. (1) The mass of baryonic matter in the disk and
bulge inside the solar radius should be larger than $3.9\times
10^{10}\, \Msun$ to produce enough microlensing events. (2) the
surface density of the dark matter inside 1.1~kpc from the plane at
the solar radius should be $30\pm 15 \, \Msun \pc^{-2}$ to satisfy the
constraints from local steller kinematics and stellar and gas content.
(3) The sum of all components should reproduce the circular velocity
curve at 2--4~kpc. Our models $A_1$ and $B_1$ easily satisfy the first
two constraints. For example, there is a mass of $4.5\times 10^{10}\,
\Msun$ of baryons inside the solar radius in our Model $A_1$, and the
dark matter surface density is $22\, \Msun \pc^{-2}$.  In our modeling
we do not use the circular velocity at 2--4~kpc because corrections
due to non-circular motions are large and uncertain. Still, even if we
accept the Binney \& Evans results for the circular velocity, the
differences appear to be small --- 10 \kms~ for 2--4~kpc, which is
within the observational uncertainty. We also note that that the
treatment of the dark matter and the bar by Binney \& Evans are not
realistic. The profile of the dark matter was assumed to be unmodified
NFW, which as we have discussed is not appropriate for the inner part
of our Galaxy where adiabatic compression will have substantially
modified the dark matter profile.  The bulge was modeled as an
elliptical exponential disk with scale-length 1~kpc. We have used more
realistic models for the dark matter and a more detailed bar model
which matches numerous observational constraints. We also rely on the
more detailed treatment of lensing events made by \citet{ZSR95} and by
\citet{Peale} for the same bar used in our analysis.  These
differences explain why we come to different conclusions.

\section{Conclusions}
\label{sec:conclusions}
We study models of the Milky Way and the Andromeda galaxies based on
the standard paradigm of disk and dark matter halo formation within a
\LCDM\ cosmology.  The models produced acceptable fits for numerous
observational data. Here we itemize our main conclusions.

(I) Cuspy NFW profiles may be compatible with observational data on MW
and M31.

(II) The compression of the dark matter by baryons is a very
significant effect in the inner parts of galaxies, and must be taken
into account. It provides a coupling between the baryons and the dark
matter. The result of this coupling is that the total density
distribution does not show any wiggles or features corresponding to
the transition from one component to another. Thus, the disk-halo
``conspiracy'' is a natural prediction of our models.

(III) The rate of microlensing events is a strong constraining factor
for models with standard cuspy halo density profiles.
These models are compatible with the recent low estimates of the
optical depth if all the stellar material inside 3.5~kpc is optimally
distributed in a $\approx 2\times 10^{10}\, \Msun$ bar.
 
(IV) In none of our models does the dark matter dominate the central
parts ($\approx 3~\kpc$) of the galaxies. Since both the MW and M31
appear to be typical high surface brightness galaxies (HSB), we expect
that this is true for other HSB galaxies. If there has been no
transfer of angular momentum between the baryons and the dark matter,
HSB galaxies should have sub-maximal disks.

(V) Including the effects of a modest transfer of angular momentum
from baryons to dark matter (expected during the non-axisymmetric
process of bar formation) can produce models with 2--3 times less dark
matter in the central $3~\kpc$ region of the Galaxy. This is likely to
allow sustanance of fast rotating bars such as those observed in the
Milky Way and M31. These models also allow a slightly heavier disk
without producing too high a rotation curve.
  
(VI) Our dynamical models suggest that the virial mass of the dark
matter halo hosting the Milky Way and M31 is in the range $10^{12}
\, \Msun < M_{\rm vir} < 2 \times 10^{12} \, \Msun$. We use the predicted
halo mass function to estimate the number density of such halos and
compare this with the observed number density of galaxies. We find
that if the K-band luminosity of the Milky Way is $M_K = -24$, as
found by \citet{DrimmelSpergel2001} based on direct fits to DIRBE
data, then the predicted number density of dark matter halos hosting
Milky Way galaxies is not inconsistent with the observed K-band
luminosity function of \citet{lf2mass}. If the Milky Way has a larger
luminosity of $M_K = -24.4$ to $~ -24.7$, as would be the case if it lies
on the Tully Fisher relation found by \citet{tp00} and
\citet{rothberg:00}, then the number density of ``Milky Way'' halos is
more than a factor of two higher than the observed number density of
galaxies at this luminosity. Increasing the virial mass of the Milky
Way's halo to the upper limit allowed by the dynamical modelling
($M_{\rm vir} < 2 \times 10^{12} \, \Msun$) allows the fainter end of the
TF range to be accomodated, but not the brightest values.

(VII) A significant fraction of the baryons within the virial radius
of the halo must not be in the disk or bulge of the Milky Way and M31.
All acceptable models required that the disk and bulge should contain
not more than $\approx$1/4-1/2 of the baryons expected to be inside
the virial radius of the halo in the absence of feedback. It is beyond
the scope of this paper to address the state and the location of the
``lost'' baryons, but this is in keeping with the usual assumptions of
semi-analytic models, and we address this issue in Paper II.  We note
that this conclusion is based on two independent arguments. Models
which have more than half of the baryons in the central luminous part
fail to produce acceptable fits for either the rotation curve or for
the local surface density. When compared with the observed luminosity
function, these models also produce too large a number density of
bright galaxies (see (VI) above). This is related to the usual
``over-cooling'' problem (see \citet{balogh:01} for a recent
discussion).

\acknowledgements

We acknowledge support from the grants NAG-5-3842 and NST-9802787.
HSZ acknowledges hospitality while visiting New Mexico State
University. AK acknowledges hospitality and support from the Institute
of Astronomy, Cambridge. We thank James Binney and Rene Walterbos for
inspiring discussions.


\end{document}